\documentclass[a4paper,11pt]{article}
\pdfoutput=1 % if your are submitting a pdflatex (i.e. if you have
\usepackage{jheppub} % for details on the use of the package, please
\usepackage[T1]{fontenc} % if needed
\usepackage{verbatim}
\usepackage{graphicx}% Include figure files
\usepackage{dcolumn}% Align table columns on decimal point
\usepackage{bm}% bold math
\usepackage{slashed}
\usepackage{braket}
\usepackage{bbold}
\usepackage{comment}

\title{\boldmath Solitons in lattice field theories via tight-binding supersymmetry}

\author[a]{Shankar Balasubramanian,}
\author[b,c]{Abu Patoary,}
\author[b,c]{and Victor Galitski}

\affiliation[a]{Center for Theoretical Physics,  Massachusetts Institute of Technology, Cambridge, MA 02139, USA}
\affiliation[b]{Joint Quantum Institute, University of Maryland, College Park, MD 20742, USA}
\affiliation[c]{Condensed Matter Theory Center, Department of Physics, University of Maryland, College Park, MD 20742, USA}

\abstract{Reflectionless potentials play an important role in constructing exact solutions to classical dynamical systems (such as the Korteweg–de Vries equation), non-perturbative solutions of various large-$N$ field theories (such as the Gross-Neveu model), and closely related solitonic solutions to the Bogoliubov-de Gennes equations in the theory of superconductivity. These solutions rely on the inverse scattering method, which reduces these seemingly unrelated problems to identifying reflectionless potentials of an auxiliary one-dimensional quantum scattering problem.  There are several ways of constructing these potentials, one of which is quantum mechanical supersymmetry (SUSY).  In this paper, motivated by recent experimental platforms, we generalize this framework to develop a theory of lattice solitons.  We first briefly review the classic inverse scattering method in the continuum limit, focusing on the Korteweg–de Vries (KdV) equation and $SU(N)$ Gross-Neveu model in the large $N$ limit.  We then generalize this methodology to lattice versions of interacting field theories.  Our analysis hinges on the use of trace identities, which are relations connecting the potential of an equation of motion to the scattering data. For a discrete Schr\"odinger operator, such trace identities had been known as far back as Toda; however, we derive a new set of identities for the discrete Dirac operator.  We then use these identities in a lattice Gross-Neveu and chiral Gross-Neveu (Nambu–Jona-Lasinio) model to show that lattice solitons correspond to reflectionless potentials associated with the discrete scattering problem.  These models are of significance as they are equivalent to a mean-field theory of a lattice superconductor.  To explicitly construct these solitons, we generalize supersymmetric quantum mechanics to tight-binding models. We show that a matrix transformation exists that maps a tight-binding model to an isospectral one which shares the same structure and scattering properties.  The corresponding soliton solutions have both modulated hopping and onsite potential, the former of which has no analogue in the continuum limit. We explicitly compute both topological and non-topological soliton solutions as well as bound state spectra in the aforementioned models.}

\begin{document} 
\maketitle
\flushbottom

\section{Introduction}
Solitons are ubiquitous in nature and occur in a plethora of classical and quantum systems, among which include water waves \cite{de-VriesPM1895, KruskalPRL1965, MiuraPRL1967, LaxCPAM1968}, optical fibers \cite{ Mollenauer_AP_2006}, magnets \cite{Kosevich_PR_1990}, superfluids \cite{Yefsah_Nature_2013, denschlag_Sci_2000, khaykovich_Sci_2002} and superconductors \cite{Efimkin_PRA_2015, Takahashi_JLTP_2014, Takahashi_PL_2012, Takahashi_PRL_2013} - an incomplete list. While solitons have been observed for centuries, the elegant mathematical structure  behind them, known as the inverse scattering method, is comparatively recent. The development of the inverse scattering method was highly influenced by the Fermi–Pasta–Ulam–Tsingou (FPUT) problem - numerical experiments carried out by Fermi et al. in 1954-1955 \cite{Fermi_osti_1955}. Fermi and co-workers were interested in the question of thermalization and sought to observe it in a discrete chain of non-linear oscillators. However, they numerically observed  a recurrent, ``almost periodic'' behavior instead, which became known as the Fermi–Pasta–Ulam–Tsingou paradox \cite{Berman_CIJNS_2005}. This paradox motivated the work by Kruskal and Zabusky ten years later, who showed that the FPUT problem maps onto the  Korteweg–de Vries (KdV) equation in the continuum limit \cite{KruskalPRL1965}. They performed their own simulations of the dynamics of the KdV equation, which showed a number of features, surprising at the time, but understood now in terms of integrable dynamics. Kruskal and Zabusky coined the term ``soliton'' and their work on the KdV equation gave impetus to the development of the inverse scattering method. The inverse scattering method eventually provided complete understanding of the nature of integrable solutions in both the KdV equations and other continuous classical integrable theories \cite{Ablowitz_SIAM_1981, Kivshar_RMP_1989, Kartashov_RMP_2011, Calogero_LNC_1976,Serkin_PRL_2007}. At the heart of the method is a mapping of the KdV equation to an auxiliary quantum scattering problem, whose time-evolved potential is a solution to the KdV equation. 

This link between the purely classical problem of waves in shallow water and quantum mechanics has led to a number of other unexpected discoveries. Notably, Dashen, Hasslacher, Neveu, Shei and others \cite{Neveu_PRD_1975, Shei_PRD_1976, Karowski_NPB_1978} showed in the 1970s that inverse scattering methods can be used to construct non-perturbative soliton solutions to saddle-point equations of $1+1$ dimensional quantum field theories (justified in the large-$N$ limit). The saddle point equations of the Gross-Neveu (GN) model are mathematically equivalent to the auxiliary scattering problem in the KdV equation. It was also shown, using trace identities for the scattering problem, that the saddle points are achieved for reflectionless potentials. 

These results further propagated to condensed matter physics. In particular, Brazovsky and others  recognized that the chiral Gross-Neveu model is equivalent to the problem  of interacting electrons in one dimension and therefore the (classical) solitons of the chiral GN model represent static solitons in either Peierels-insulating charge-density-wave states (for repulsive fermions) or superconducting states (for attractive fermions) \cite{brazovskii_JETPL_1980,brazovskii_JETPL_1981,SSH, Campbell_PRB_1981,buzdin_SPJETP_1983,horovitz_PRL_1981,Gerald_PRL_2008,Takahashi_PRL_2013, Takahashi_PL_2012, Takahashi_JLTP_2014}. In these cases too, the solitons originated from the reflectionless self-consistent solutions of the Bogoliubov-de Gennes equations -- the classical saddle point of the action describing the superconductor/Peierels insulator. More recently, these methods were applied to Larkin-Ovchinnikov-Fulde-Ferrel superconducting phases \cite{larkin_SPJETP_1965, Flude_PR_1964} and time-dependent superconductors, featuring a solitonic lattice and periodic-in-time soliton trains of the superconducting order parameter correspondingly \cite{Gerald_PRL_2008, Barankov_PRL_2004, Yuzbashyan_PRB_2008, buzdin_SPJETP_1987, Basar_PRD_2008, Basar_PRD_2009, Correa_AP_2009, Efimkin_PRA_2015,Tanaka_PRL_2001, lin_NJP_2012,garaud_PRL_2011}. Moving solitons have also been considered in both high-energy context \cite{weinberg_CMMP_2012} and for cold-atom superfluids \cite{Yefsah_Nature_2013, Burger_PRL_1999, denschlag_Sci_2000, khaykovich_Sci_2002, strecker_Nat_2002, Kartashov_RMP_2011}. Another more controversial example has been the attempt to construct soliton trains in imaginary time (to represent hypothetical at this stage imaginary-time crystals) \cite{Galitski_PRB_2010, Wilczek, Efetov}.  In all these cases however, the mathematical construction of solitons hinges on the reflectionless potentials of the underlying scattering problem. 

A useful method for constructing such potentials is supersymmetric quantum mechanics (for a comprehensive review, see \cite{COOPER1995267}). Continuous quantum-mechanical supersymmetry involves writing the Schr{\"o}dinger operator - the Hamiltonian - as a product of two operators $H=A^\dagger A$ and constructing another Schr{\"o}dinger operator $\widetilde H =A A^\dagger$. It can be shown that the two corresponding Hamiltonians/potentials are (almost) isospectral and  share the same scattering data. This allows construction of non-trivial reflectionless potentials. The simplest example is to start with the obviously transparent constant potential and find its superpartner, that happens to be related to the simplest KdV single-soliton solution. The process can be iterated and a series of reflectionless potentials can be constructed including multi-soliton solutions. 

All in all, the combination of the inverse scattering method and quantum-mechanical supersymmetry (SUSY) allows to build exact analytic non-linear soliton solutions to a variety of diverse problems from bound states in large-$N$ field theories to non-linear excitations in superfluids and superconductors, with both static and dynamic solitons recently of interest. 

The resolution of the Fermi-Pasta-Ulam ``paradox'' has been understood through both KdV limit and the Toda lattice -- a fully integrable classical lattice model, which gives rise to the FPUT problem as its Taylor expansion \cite{Toda_Springer_1989}. However, surprisingly the lattice field theoretical and superconducting analogies of the Toda lattice have not been fully explored. A large number of integrable lattice models have  been considered using more modern methods in quantum integrability that go beyond classical inverse scattering \cite{korepin_CUP_1997, faddeevbethe}. However, it appears that the role of reflectionless potentials and their explicit construction have not been  fully investigated in the context of interacting tight-binding models. This paper fills this gap and develops a general framework to study solitons in a family of large-$N$ lattice field theories. 

A key new step, introduced here, is the generalization of quantum mechanical SUSY to a wide class of tight-binding models (which we refer to as discrete SUSY). The existence of such a SUSY transform is not obvious, because the continuous SUSY transformation does not modify the kinetic energy operator - the Laplacian in the usual Schr{\"o}dinger equation - while tight-binding models contain a hopping dependent ``kinetic energy''.  Still, we show that local tight-binding models in a potential map under discrete SUSY onto the same class of models, but with generally site-dependent hoppings.  The scattering data are preserved under discrete SUSY, which is useful in order to generate a series of reflectionless potentials.  For lattice versions of the Gross-Neveu and chiral Gross-Neveu models (equivalent to lattice superconductors), we rewrite the action in terms of scattering data and show that saddle point solutions in the large-$N$ limit correspond to reflectionless potentials.  We construct the bound state spectra for both models, and utilize discrete SUSY to explicitly compute both topological and non-topological soliton solutions.  

\section{Gross-Neveu model and Korteweg-de Vries equation}
We start with a review of soliton solutions in continuum systems.  The simplest model which exhibits such solutions is the Korteweg-de Vries equation, which was first used to describe shallow water waves \cite{de-VriesPM1895}.  The KdV equation can be generalized to construct a hierarchy of nonlinear differential equations, each of which also supports soliton solutions.  A seemingly unrelated system is the Gross-Neveu model, which is field theory of $N$ flavors of massless fermions subject to an attractive potential.  When $N$ is large, the Gross-Neveu model is a toy model of quantum chromodynamics: it is asymptotically free, and spontaneously breaks chiral symmetry, which results in the fermions acquiring a mass, therefore gapping the theory.  The mean field theory of a chiral version of the Gross-Neveu model, also known as the Nambu-Jona-Lasinio model, can be shown to be equivalent to BCS superconductivity in 1D \cite{Gerald_PRL_2008, Takahashi_PRL_2013}.  Both the Gross-Neveu and the chiral Gross-Neveu model have nontrivial saddle point solutions which correspond to solitons.

\begin{figure}
    \centering
    \includegraphics[width=\textwidth]{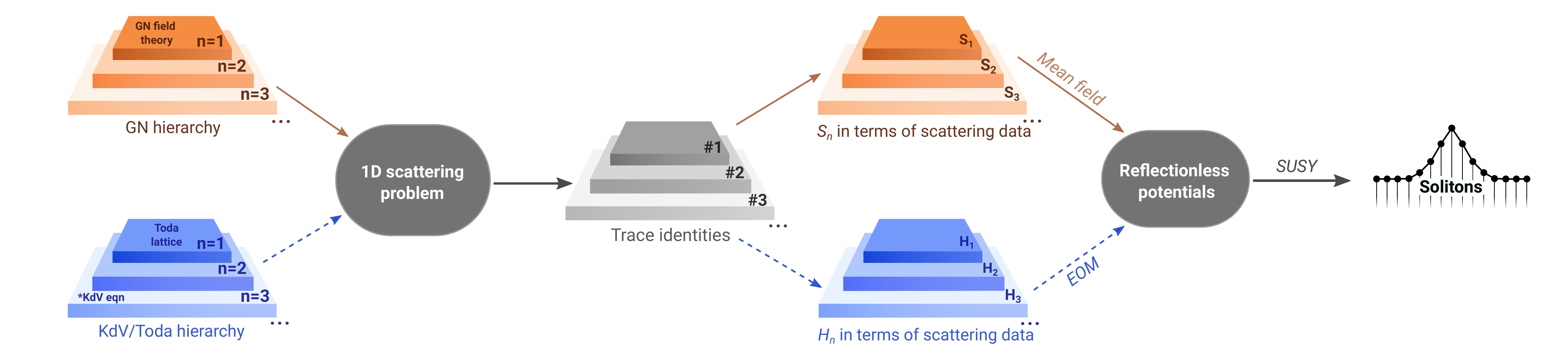}
    \caption{A schematic depicting the sequence of steps used to generate a soliton hierarchy.  We start with a hierarchy of field theories (the first of which is the Gross-Neveu field theory) or a hierarchy of dynamical systems/partial differential equations (the first in the Toda hierarchy is the Toda lattice, and the third in the KdV hierarchy is the KdV equation).  The equations of motion of the GN field theory and a suitable change of variables for the KdV equation/Toda lattice map these seemingly unrelated systems to a one-dimensional scattering problem.  With the help of trace identities derived in this scattering problem, we may rewrite either the action or the Hamiltonian in terms of scattering data, i.e. the reflection coefficient, the transmission coefficient, and the bound state energies of the scattering problem.  There are also a hierarchy of trace identities, and each one uniquely maps onto one of the field theories or dynamical systems in the original hierarchies.  The saddle point of the action corresponds to a reflectionless potential, and the equations of motion of the Hamiltonian systems imply the conservation of the magnitude of the reflection coefficient.  In both cases, reflectionless potentials are either valid saddle points of the action or solutions to the dynamical system.  These potentials are retrieved explicitly using supersymmetry methods, or more elaborately using the Gelfand-Levitan equation and the inverse scattering formalism.}
    \label{fig:continuousSchematics}
\end{figure}

It turns out that the construction of soliton solutions for the KdV equation and the Gross-Neveu field theories are nearly identical.  This is because both the KdV equation as well as the equations of motion for the Gross-Neveu model can be mapped onto a 1D quantum scattering problem.  Based on a series of \emph{trace identities} that connect the potential of the scattering problem to the scattering properties of plane wave solutions, it can be shown that soliton solutions correspond to reflectionless potentials of the scattering problem.  Constructing these reflectionless potentials can be achieved through several methods, one of which is quantum mechanical supersymmetry.  In this section, we will briefly review these results, which have been summarized in Figure~\ref{fig:continuousSchematics}.  
\subsection{Gross-Neveu (GN) model}
The Lagrangian for the Gross-Neveu model in $(1+1)$ dimensions is given by \cite{Neveu_PRD_1975}:
\begin{equation}
    \mathcal{L} = \sum_{k=1}^{N}i\overline{\psi}^{(k)}\slashed{\partial}\psi^{(k)} +\frac{g^2}{2}\left(\sum_{k=1}^{N}\overline{\psi}^{(k)}\psi^{(k)}\right)^2, \label{GN without HS}
\end{equation}
where $\psi^{(k)}(x,t)$ is a two-component fermionic field and $k$, which ranges from $1$ to $N$, labels fermion flavor.  Throughout the analysis, we assume that $N$ is large, which justifies the use of mean field theory.  We also adopt the standard definitions $\overline{\psi}^{(k)}=\psi^{\dagger {(k)}}\gamma_0$ and $\slashed{\partial} = \gamma_{\mu}\partial^{\mu}$ with $\gamma_{\mu}$ being the Dirac matrices in (1+1) dimension.  Throughout the text, we will choose the representation where $\gamma_0 = -\sigma_x$, $\gamma_1 = -i \sigma_z$, and $\gamma_5 = \gamma_0 \gamma_1 = \sigma_y$.  For the sake of brevity, we also introduce the following notations:
\begin{align}
    \overline{\psi}\slashed{\partial}\psi & \equiv \sum_{k=1}^{N}\overline{\psi}^{(k)}\slashed{\partial}\psi^{(k)}, \\
    \overline{\psi}\psi & \equiv \sum_{k=1}^{N}\overline{\psi}^{(k)}\psi^{(k)}.
\end{align}
In the path integral formulation the real time functional can be written as
\begin{equation}
Z = \int \mathcal{D}\psi \mathcal{D}\overline{\psi} \exp\left[i\int d^2x\,\left(-i\overline{\psi}\slashed{\partial}\psi-\frac{g^2}{2}\left(\overline{\psi}\psi\right)^2\right)\right]
\end{equation}
We introduce a Hubbard-Stratonovich field $\Delta(x,t)$, so that the path integral becomes
\begin{equation} \label{GN with HS}
   Z = \int \mathcal{D}\psi \mathcal{D}\overline{\psi}\mathcal{D}\Delta \exp\left[i\int d^2x\,\left(-i\overline{\psi}\slashed{\partial}\psi-g\Delta\overline{\psi}\psi-\frac{\Delta^2}{2}\right)\right],
\end{equation}
 Following a standard mean field assumption, we assume that the Hubbard-Stratonovich field possesses no dynamics, and we search for static solutions which correspond to the saddle point of the action.  We further restrict our search of saddle point solutions to those that satisfy $\lim_{|x|\rightarrow \infty}\Delta(x) = \Delta_0$. Due to the time independence of the HS field, the equations of motion for the fermionic fields map onto a one-dimensional scattering problem for each component of the fermionic field; the effective scattering problem is equivalent to the time independent Schr{\"o}dinger equation
\begin{equation} \label{GN Schrod}
    -\partial_x^2 \psi_{1,2} + u_{1,2}(x) \psi_{1,2} = k^2 \psi.
\end{equation}
where $\psi_{1}$ and $\psi_{2}$ are two components of the fermion, $k^2$ is the energy and the effective potential
\begin{equation} \label{u as HS field}
    u_{1,2}(x) \triangleq u_\pm(x) = g^2(\Delta^2-\Delta_0^2) \pm g\frac{d\Delta}{dx}
\end{equation}
 vanishes asymptotically. The saddle point solutions can therefore be constructed by recasting the action in terms of scattering data of this scattering problem.
 \subsection{Korteweg-de Vries (KdV) equation} \label{KdV}
 The KdV equation is a non-linear partial differential equation used to describe a wide range of phenomena from waves on a shallow water surface \cite{de-VriesPM1895, Crighton1995, segur_JFM_1973} to ion-acoustic waves in plasma \cite{Herskowitz_PRL_1972, Hershkowitz_PRL_1974}. For a function $u(x,t)$ the KdV equation is
 \begin{equation} \label{KdV equation}
     \frac{\partial u}{\partial t} - 6u\frac{\partial u}{\partial x} + \frac{\partial^3 u}{\partial x^3} = 0.
 \end{equation}
 At first sight, Eqn. \eqref{KdV equation} does not seem to have anything in common with Gross-Neveu model discussed in the previous subsection. However, the KdV equation can also be mapped onto a time independent Schr{\"o}dinger equation, whose effective potential coincides with the function $u(x,t)$. To explicitly see this mapping, define a function $v(x,t)$ which satisfies
 \begin{equation} \label{Defn of v}
     u = v^2 + \frac{\partial v}{\partial x}.
 \end{equation}
 Substituting $v = \frac{\partial \log{\psi}}{\partial x}$ into Eqn. \eqref{Defn of v} one finds
 \begin{equation} \label{u as a fn of psi}
     u = \frac{ \partial_x^2 \psi}{\psi}.
 \end{equation}
 Note that if $u(x,t)$ is a solution of KdV equation then so is $\Bar{u}(x,t) = u(x+6\lambda t, t) + \lambda$.  This foreshadows the fact that the KdV equation possesses soliton solutions which move without changing shape. Thus, we can replace $u$ in Eqn. \eqref{u as a fn of psi} with $u-\lambda$ to write it as
 \begin{equation} \label{KdV Schrodinger}
      -\partial_x^2 \psi + u\psi = \lambda \psi.
 \end{equation}
 If we set $\lambda = k^2$ then Eqn. \eqref{KdV Schrodinger} is same as Eqn. \eqref{GN Schrod} derived for the GN model.  In the KdV equation, propagating solitons are described by the time-evolved potential $u(x,t)$; under the mean field assumption, solitons are static in the GN model and are described by $\Delta(x)$, which is related to $u(x,t=0)$ via Eqn.~\eqref{u as HS field}.
 \subsection{Trace identities for Schr{\"o}dinger equation} \label{Trace Ident Cont}
 Both the GN model and the KdV equation map onto a time independent Schr{\"o}dinger equation. Solutions to the Schr{\"o}dinger equation can be both bound as well as scattering, and the scattering data must account for both types of states. In the asymptotic limit $x \to \pm \infty$, scattering states have the form
\begin{align}
    \lim_{x\to -\infty} \psi(x,k) &= e^{ikx} + r(k)e^{-ikx} \label{Scattering cont -Infty}\\
    \lim_{x\to \infty} \psi(x,k) &= t(k)e^{ikx}, \label{Scattering cont Infty}
\end{align}
where $r(k)$ and $t(k)$ are the momentum-dependent reflection and transmission coefficients respectively. For a bound state with energy $-\kappa_{\ell}^2$, the wavefunction has the asymptotic form
\begin{equation}
    \lim_{|x|\to \infty} \psi_{\ell}(x) = N_{\ell} e^{-\kappa_{\ell}|x|},
    \label{Bound cont}
\end{equation}
where $N_{\ell}$ is a normalization. We now define the scattering data as the set $\left\{r(k), t(k), \kappa_\ell\right\}$. Remarkably, there exists a hierarchy of \emph{trace identities} that relates integrals of powers and derivatives of the potential to the reflection coefficient and the bound state energies \cite{Faddeev_FAA_1971,Zakharov_TMP_1974, Takhtadzhyan_TMP_1974, Deift_CPAM_1979}. Suppose the potential we are given has $N$ bound states.  Then, the trace identities are given by
\begin{equation}
    \frac{1}{2\pi i}\int_{-\infty}^{\infty}q^{2n}\log\left[1-|r\left(q\right)|^2\right]\,dq -\frac{2}{2n+1}\sum_{\ell=1}^{N}(i\kappa_\ell)^{2n+1} = -\frac{1}{(2i)^{2n+1}}\int_{-\infty}^{\infty}\beta_{2n+1}(x)\,dx,
    \label{Trace Identities Cont}
\end{equation}
where $n$ is a non-negative integer and $\beta_n$ are functions of the potential which are recursively determined from the coupled differential equations
\begin{align}
    \beta_{n+1}(x) &= -\frac{d}{dx}\beta_n(x) - \sum_{m=1}^{n-1}\beta_m(x)\beta_{n-m}(x) \\
    \beta_1(x) &= u(x).
\end{align}
For clarity and convenience, we explicitly reproduce the first three identities below:
\begin{align} 
    -\frac{1}{2i}\int_{-\infty}^{\infty} u\,dx &= \frac{1}{2\pi i}\int_{-\infty}^{\infty}\log {\left(1-|r\left(q\right)|^2\right)}\,dq -2\sum_\ell(i\kappa_\ell), \label{n=1 identity} \\
    -\frac{1}{8i}\int_{-\infty}^{\infty} u^2\,dx &= \frac{1}{2\pi i}\int_{-\infty}^{\infty}q^2\log {\left(1-|r\left(q\right)|^2\right)}\,dq -\frac{2}{3}\sum_\ell(i\kappa_\ell)^3, \label{n=2 identity} \\
    -\frac{1}{32i}\int_{-\infty}^{\infty} (2u^3 + u_x^2)\,dx &= \frac{1}{2\pi i}\int_{-\infty}^{\infty}q^4\log {\left(1-|r\left(q\right)|^2\right)}\,dq -\frac{2}{5}\sum_\ell(i\kappa_\ell)^5. \label{n=3 identity} 
\end{align}
In particular, the first identity relates the integral of the potential to the reflection coefficient and bound state energies; this will be useful for performing a change of variables in the Gross-Neveu action.
\subsection{Soliton solutions are reflectionless potentials} \label{Saddle point}
One can formulate both the Gross-Neveu mean field action and the KdV equation in terms of scattering data using the trace identities of section \ref{Trace Ident Cont}. For the GN model, the Lagrangian in Eqn. \eqref{GN with HS} is quadratic in the fermion fields $\psi$ and $\overline{\psi}$. Consequently the fermions can be integrated out, yielding an effective action for the HS field
\begin{equation}
    S_{\text{eff}}(\Delta) = -\log{\left[\int \mathcal{D}\psi \mathcal{D}\overline{\psi} \exp\left(-\int_{-\infty}^{\infty} dx\int_{0}^{T} dt\,\overline{\psi}\left(i\slashed{\partial}-g\Delta\right)\psi\right)\right]}-\frac{1}{2}\int_{-\infty}^{\infty} dx\int_{0}^{T} dt\, \Delta^2. \label{Seff as HS field}
\end{equation}
Using the first trace identity given by Eqn. \eqref{n=1 identity} and other standard results from path integration and scattering theory one can write the effective action as a function of scattering data \cite{Neveu_PRD_1975} as
\begin{multline} \label{GN action Scattering data}
    S_{\text{eff}} \propto \frac{1}{g^2} \int_{-\infty}^{\infty} \log{\left[1-|r\left(k\right)|^2\right]}dk  +\frac{4\pi}{g^2}k_0 \\ -\frac{1}{\pi} \int_{0}^{\infty}\frac{k dk}{\sqrt{k^2+g^2\sigma_0^2}}\left(\log{\left(1-|r\left(k\right)|^2\right)}+4\pi \tan^{-1}\frac{k_0}{k} \right) + \text{const.}
\end{multline}
where it has been assumed that there is only one bound state with energy $-k_0^2$. The saddle point of the effective action can be determined by first extremizing over the reflection coefficient; we see that the minimum corresponds to a reflectionless potential, i.e the choice of $\Delta(x)$ such that the reflection coefficient $r(k)=0$.

A similar connection to reflectionless potentials can be made for the KdV equation. It has been has shown that the KdV equation is a Hamiltonian system and arises from the Hamiltonian $H[u] = \int_{-\infty}^{\infty} dx \left[u^3(x)+\frac{u_x^2}{2}\right]$ \cite{Gardner_JMP_1971}. One can check that the Hamiltonian $H$ yields the KdV equation in the form
\begin{equation} \label{KdV Eqn from Hamiltonian}
    \frac{\partial u}{\partial t} = \frac{\partial}{\partial x} \frac{\delta H}{\delta u},
\end{equation}
where $\frac{\delta H}{\delta u}$ indicates a functional derivative of $H$. Using the third non-trivial trace identity (Eqn. \eqref{n=3 identity}), this Hamiltonian can be written solely in terms of scattering data as
\begin{equation} \label{KdV Hamiltonian in terms of scattering data}
    H_{\text{KdV}} = -\frac{8}{\pi}\int_{-\infty}^{\infty}k^4\log{\left(1-|r(k)|^2\right)dk -\frac{32}{5}\sum_{i=1}^N k_i^5},
\end{equation}
where $N$ is the number of bound states and $-k_i^2$ is the energy of the $i$th bound state. However, one must argue that the scattering data can be rewritten as pairs of canonically conjugate variables.  This was first verified by Faddeev and Zakharov, and the corresponding Hamilton's equations of motion imply that one class of solutions to the KdV equation are time-dependent reflectionless potentials.  In fact, a particular combination of the scattering data turns out to be action-angle coordinates for the KdV Hamiltonian, which manifestly indicates the integrable nature of the system (see \cite{Faddeev_FAA_1971, Faddeev_2007_Springer}).

The remaining question left is how to construct such potentials, which correspond to soliton solutions.  One such method utilizes supersymmetric quantum mechanics, which is briefly summarized in section \ref{Cont SUSY}. In particular, applying the supersymmetry transformation to a constant potential results in a nontrivial reflectionless potential for the Schr\"odinger equation; this corresponds to a kink soliton.

Note that for the KdV equation the third trace identity serves as the Hamiltonian. However, in general, one can consider any of the trace identities as a candidate Hamiltonian and construct equations of motion; each such equation of motion corresponds to a nonlinear higher-order differential equation for $u(x,t)$, each of which is integrable. The set of equations constructed in this manner form the KdV hierarchy \cite{LaxCPAM1968, drazin_johnson_1989}.

\subsection{Field theory hierarchy}
While the KdV hierarchy is a celebrated concept in classical integrability theory, it is possible to construct a similar hierarchy for the Gross-Neveu field theories.  Note that only the first trace identity given by Eqn. \eqref{n=1 identity} has been used to write the action of GN model as a function of the scattering data. We may similarly use the other trace identities to construct a hierarchy of field theories whose saddle point corresponds to reflectionless potential. To construct the field theory after the Gross-Neveu model in the hierarchy, we replace the first trace identity by hand with the second trace identity, obtaining the action
\begin{equation} \label{Lagranian second cont field theory}
    \mathcal{L}_2 = i\overline{\psi}\slashed{\partial}\psi -g\Delta\overline{\psi}\psi-\left(g^4\Delta^4+g^2\Delta \frac{\partial^2 \Delta}{\partial x^2}\right).
\end{equation}
One can integrate out the HS field to write the Lagrangian only in terms of the fermions as
\begin{equation}
    \mathcal{L}_2 = i\overline{\psi}\slashed{\partial}\psi - \log\left(\int d\Delta \exp{\left[\int dx\left( ig\Delta\overline{\psi}\psi-g^2\Delta\frac{\partial^2 \Delta}{\partial x^2}-g^4\Delta^4\right)\right]}\right).
\end{equation}
This is the second field theory in the hierarchy. In principle, one can construct one such field theory per trace identity; however, each higher order field theory is intractable and non-local when one attempts to integrate out the HS field.

\section{Solitons in a lattice Gross-Neveu model}
In the continuum Gross-Neveu field theory, it was argued that stationary solutions to the action correspond to choices of $\Delta(x)$ such that the equations of motion describe a particle in a reflectionless potential.  In lattice systems, which are currently achievable in cold atom experiments, there are several differences that may result in exotic soliton solutions.  Firstly, there exist non-unique discretization schemes of placing fermions on a lattice that give rise to different microscopic dynamics even if they reduce to the same theory in the continuum limit.  Secondly, if a lattice theory possesses soliton solutions, then a solitonic texture may arise in both the onsite potential and the hopping, which is not seen in the continuum limit.  And lastly, at the scale of the lattice spacing, the dispersion relation looks like a band structure, thus differing markedly from the dispersion in the continuum limit; this may result in different characteristic soliton spectra and propagation velocities (the latter of which we do not consider in this paper).  The aforementioned reasons therefore make it enticing to study solitons in discretized field theories.

For lattice systems we will adopt a reverse engineering approach, starting with known discrete trace identities due to Toda \cite{Toda_Springer_1989} and using these identities to subsequently construct lattice models whose saddle points correspond to reflectionless potentials.  In Section \ref{chiralGNsection}, we perform the same analysis for the chiral Gross-Neveu model which is an effective description for a lattice superconductor.  In the process, we derive previously unknown trace identities for the discrete Dirac operator.

\subsection{Trace identities for tight-binding Hamiltonians}
Before introducing the lattice Gross-Neveu model, we first set out to derive trace identities for discrete scattering problems. To define the scattering problem we consider a generic tight binding Hamiltonian with nearest neighbour hopping, written out explicitly as
\begin{equation}\label{discreteschroham}
    \mathcal{H} = \sum_{n = -\infty}^\infty \left(-t_{n+1} |n\rangle \langle n+1| + \text{h.c.} + u_n |n\rangle \langle n|\right),
\end{equation}
The Hamiltonian describes a particle in a discrete Schr\"odinger equation, which we write as 
\begin{equation} \label{Schrod: General Discrete}
    -(t_{n+1}\psi_{n+1}+ t_{n}\psi_{n-1}) + u_n\psi_n = \lambda \psi_n,
\end{equation}
where $n$ is the site number, $t_{n}$ is a site-dependent hopping which is absent in the continuum case, and $u_{n}$ is a discrete onsite potential.  Both $t_{n}$ and $u_{n}$ will later be shown to be expressible in terms of the HS parameter. The energy $\lambda$ is independent of $n$, and we can determine it by constructing eigenfunctions in the asymptotic limit when $n \to \infty$; these eigenfunctions satisfy
\begin{equation} \label{Asymptot Schrod: Discrete}
    -(\psi_{n+1}+\psi_{n-1}) + u \psi_n = \lambda \psi_n,
\end{equation}
where we have used that $\lim_{n\to \pm \infty} u_n = u$ and $\lim_{n\to \pm \infty} t_n = 1$. The solution for Eqn. \eqref{Asymptot Schrod: Discrete} is $\psi_{n}=e^{i\phi n}\equiv z^{n}$ where $\phi$ is a quasi-momentum constrained to lie within the interval $[0,2\pi)$; this gives the dispersion $\lambda = 2\cos{\phi} + u$.

Next, we will reproduce the trace identities analogous to Eqn. \eqref{Trace Identities Cont} that relate the potential $u_{n}$ to the scattering data of the discrete Schr{\"o}dinger equation, which were previously known to Toda \textit{et al.} \cite{Toda_Springer_1989, Flaschka_1974_PTP}. We first make a choice of two independent basis functions of Eqn. \eqref{Schrod: Discrete}. We denote those two bases as $\{f_{n}(z),f_{n}(\frac{1}{z})\}$ and $\{g_{n}(z),g_{n}(\frac{1}{z})\}$; their asymptotic behavior is given by
\begin{align}
   \lim_{n\to\infty} f_n(z) &= z^n, \\
   \lim_{n\to -\infty} g_n(z) &= z^{-n}.
\end{align}
These are known as the \emph{Jost functions}, which is a frequently used representation in scattering theory.  This choice of basis is necessary in order to derive well-behaved trace identities. Since both $f_{n}$ and $g_{n}$ independently form a basis, one may be written as a linear combination of the other. Thus, we have the relation
\begin{align}
    f_n(z) &= b(z) g_n(z) + a(z)g_n\left(\frac{1}{z}\right), \label{a,b coefficients 1}
\end{align}
where we have introduced coefficients $a$ and $b$ which can be shown to be related to the reflection and transmission coefficients of the original discrete scattering problem. A valid scattering eigenstate for the discrete Schr{\"o}dinger equation in terms of the Jost functions is $\psi_{n}(z)=\frac{f_{n}(z)}{a(z)}$; the asymptotic behavior of this solution assumes a form familiar from scattering theory:
\begin{equation}
    \psi_{n}(z) =
    \begin{cases}
    z^n + \frac{b(z)}{a(z)}z^{-n}, & \text{for } n \to -\infty 
    \\[10pt]
    \frac{1}{a(z)}z^n, & \text{for } n \to \infty
  \end{cases} \label{Asymptotic psi}
\end{equation}
From Eqn. \eqref{Asymptotic psi} we find that the reflection coefficient $r(z)$ and the transmission coefficient $t(z)$ are related to $a(z)$ and $b(z)$ according to the relations $r(z)=\frac{b(z)}{a(z)}$ and $t(z)=\frac{1}{a(z)}$. Since the reflection and transmission coefficients satisfy $|r|^2+|t|^2=1$ we have
\begin{equation}
    |a(z)|^2 = \frac{1}{1-|r(z)|^2}. \label{r and a relation}
\end{equation}
So far, we have considered the scattering states for the discrete Schr\"odinger operator; it may consist of discrete bound states as well, which are necessary for a complete characterization of the scattering data.  The location of the bound states occur at $z=z_j$ for which $a(z_j)=0$; these correspond to eigenstates which scale as $z_j^n$ for large $n$.  We also require the physical constraint that $|z_j|<1$ in order for the wavefunction to be normalizable and decay exponentially at infinity.  Thus, we require the following constraints on $a(z)$:
\begin{align}
    |a(z)|^2 &= \frac{1}{1-|r(z)|^2} \text{  when } |z|=1, \label{scattering a}\\
    a(z_j) &= 0 \text{     for some discrete } |z_j|<1 .\label{bound a}
\end{align}
Appealing to the theory of meromorphic functions, the properties of $a(z)$ allow us to uniquely define $\log |a(z)|$ within the unit disk $|z|<1$ via the Poisson-Jensen theorem:
\begin{equation} \label{ln(a) and scattering data: Discrete}
    \log{|a(z)|} = \sum_{j=1}^{M}\log {\left|\frac{z-z_j}{1-z_jz}\right|}-\frac{1}{4\pi} \int_{0}^{2\pi}\Re\left( \frac{e^{i\phi}+z}{e^{i\phi}-z}\right)\log{\left[1-|r(e^{i\phi})|^2\right ]}\,d\phi,
\end{equation}
where $M$ is the number of bound states and $\Re[z]$ means the real part of $z$. We can expand the right hand side of the equation above using the following series expansions:
\begin{align}
\label{power series 1}
    \log{\frac{z-z_j}{1-z_jz}} &= \log{z_j} + \sum_{p=1}^{\infty}\frac{z_j^p-z_j^{-p}}{p}z^p,\\
\label{power series 2}
    \Re\left( \frac{e^{i\phi}+z}{e^{i\phi}-z}\right) &= 1 + 2\sum_{p=1}^{\infty}\cos{(p\phi-p\theta)}|z|^p.
\end{align}
where $z=|z|e^{i\theta}$. Substituting these expansions, we may write
\begin{equation} \label{ln(a) power series of scattering data}
    \log \frac{|a(z)|}{|a(0)|} = \sum_{p=1}^{\infty}\left[\sum_{j=1}^{M}\Re\left(\frac{z_j^p-z_j^{-p}}{p}z^p\right) -\frac{|z|^{p}}{2\pi} \int_{0}^{2\pi}\log{\left[1-|r(e^{i\phi})|^2)\right]}\cos{(p\phi-p\theta)}d\phi \right].
\end{equation}
One can write $\log(a(z)/a(0))$ to the onsite potential via the following relation: 
\begin{equation} \label{connceting a with potential}
    -\log{\frac{a(z)}{a(0)}} = \sum_{n=1}^{\infty}\frac{(-1)^n\text{Tr}\left(H^n-(H_0)^n\right)}{n(z+\frac{1}{z})^n},
\end{equation}
where $H$ and $H_{0}$ are tridiagonal matrices; $H$ is the Hamiltonian for the Schr\"odinger operator with elements $H_{n,n-1}=-t_{n}, H_{n,n+1}=-t_{n+1}$, and $H_{n,n} = u_n$, and $H_0$ is the asymptotic form of the Hamiltonian with matrix elements $(H_0)_{n,n-1} =(H_0)_{n,n+1} =-1$  and $(H_0)_{n,n} = u$. The derivation of the Eqn. \eqref{connceting a with potential} is given in Appendix \ref{Appendix A}.  In section \ref{chiralGNsection}, we derive similar identities for the Dirac operator in the case of the chiral Gross Neveu model.

We can expand the right hand side of Eqn. \eqref{connceting a with potential} as a power series of $z$ to write
\begin{equation} \label{Kp equation }
   \sum_{p=1}^{\infty} K_pz^p = \sum_{n=1}^{\infty}\frac{(-1)^{n+1}\text{Tr}\left(H^n-(H_0)^n\right)}{n(z+\frac{1}{z})^n},
\end{equation}
where $K_p$ are real coefficients that depend on the potential and the hopping. 
We have already derived an alternate expression for $\log{(|a(z)|/|a(0)|)}$ in Eqn. \eqref{ln(a) power series of scattering data} which we can compare with Eqn. \eqref{Kp equation } to write
\begin{equation}
    \label{Trace identity for complex z}
    \sum_{p=1}^{\infty} K_{p}|z|^{p}\cos{p\theta} = \sum_{p=1}^{\infty}\Bigg[-\frac{|z|^{p}}{2\pi} \int_{0}^{2\pi}\log{\left(1-|r(e^{i\phi})|^2\right)}\cos{(p\phi-p\theta)}\,d\phi +\sum_{j=1}^{M}\Re\left(\frac{z_j^p-z_j^{-p}}{p}z^p\right)\Bigg].
\end{equation} 
%Note that if $z$ has non-zero imaginary part then from Eqn. \eqref{Trace identity for complex z} will impose potential independent constraints on scattering data which is unphysical. 
In general, and for the cases we will consider, the Hamiltonian is real; thus the eigenstates can be chosen to be real and we expect the bound states to monotonically decay at infinity without spatial oscillations.  This implies that $z_j$ is real.  With this simplification, we set $\theta = 0$ and compare the coefficients of $z^p$ in the above equation to get
\begin{equation} \label{Trace Identities}
      K_{p} = -\frac{1}{2\pi} \int_{0}^{2\pi}\log{\left(1-|r(e^{i\phi})|^2\right)}\cos{(p\phi)}\,d\phi +\sum_{j=1}^{M}\frac{z_j^p-z_j^{-p}}{p}.
\end{equation}
Matching to the series expansion in terms of the onsite potential and hopping yields a discrete version of the trace identities discussed in Dashen et al. \cite{Neveu_PRD_1975}.  For convenience, we write the first two identities associated with matching the two lowest order terms in both series expansions:
\begin{align}
\sum_{n=-\infty}^{\infty} u_{n} &= -\frac{1}{2\pi} \int_{0}^{2\pi}\log{\left(1-|r(e^{i\phi})|^2\right)}\cos{(\phi)}\,d\phi +\sum_{j=1}^{M}\left(z_j-\frac{1}{z_j}\right), \label{Discrete Trace Id 1} \\
\sum_{n=-\infty}^{\infty} u_{n}^2+2t_n^2 &= -\frac{1}{2\pi} \int_{0}^{2\pi}\log{\left(1-|r(e^{i\phi})|^2\right)}\cos{(2\phi)}\,d\phi +\frac{1}{2}\sum_{j=1}^{M}\left(z_j^2-\frac{1}{z_j^2}\right).
\end{align}
Much like in the continuous case, sums of the hopping and onsite potentials can be related directly to the reflection coefficient of the scattering problem and the locations of the bound states.
\subsection{Lattice Gross-Neveu model}

Motivated by the continuous case we can formulate the discrete version of the Gross-Neveu model. For a discrete two-component fermionic field $\Psi_{n}(t)$, where $n$ is the lattice site index, the action $S = \sum_n \mathcal{L}_n$ can be specified by the lattice Lagrangian
\begin{equation}
    \mathcal{L}_n =  \sum_{k=1}^{N}\Psi_{n}^{\dagger(k)}(t)\begin{pmatrix} i\frac{\partial}{\partial t} & \partial_{n} \\ \partial_{n}^{\dagger} & i\frac{\partial}{\partial t} \end{pmatrix} \Psi_{n}^{(k)}(t)
    +\frac{g^2}{2}\left(\Psi_{n}^{\dagger (k)}(t)\gamma_0\Psi_{n}^{(k)}(t)\right)^2, \label{GN discrete}
\end{equation}
%\begin{pmatrix}\psi_{n}^{(1)\dagger} & \psi_{n}^{(2)\dagger}  \end{pmatrix}
%\begin{pmatrix} \psi_{n}^{(1)} \\ \psi_{n}^{(2)} \end{pmatrix}
where $N$ is the number of flavors of fermions and $\Psi_{n}^{(k)}(t) = \left(\psi_{n}^{(1,k)}(t),\psi_{n}^{(2,k)}(t)\right)^{T}$ can be written as a two-component field.  Throughout the text, we will assume that $N$ is large in order to justify subsequent saddle point computations. The operators $\partial_n$ and $\partial_{n}^{\dagger}$ the analogues of derivative operators in the continuum limit but will be represented as discrete ``shift'' operators with matrix elements $\partial_n f_{n} \triangleq \tau_{n+1}f_{n+1}$. The fermionic field is always assumed to be time dependent but for the sake of brevity we will not explicitly write it. We also suppress the sum over flavors by introducing the following notation:
\begin{equation}
    \Psi_{n}^{\dagger}\mathbf{M}\Psi_{n} \equiv \sum_{k=1}^{N} \Psi_{n}^{\dagger(k)}\mathbf{M}\Psi_{n}^{(k)},
\end{equation}
where $\mathbf{M}$ is a $2 \times 2$ matrix. We introduce a discrete Hubbard-Stratonovich (HS) field $\Delta_n(t)$ to write an equivalent Lagrangian:
\begin{equation}
    \mathcal{L}_{n} = \Psi_{n}^{\dagger} \begin{pmatrix} i\frac{\partial}{\partial t} & \partial_{n}+g\Delta_n(t) \\ \partial_{n}^{\dagger}+g\Delta_{n}(t) & i\frac{\partial}{\partial t} \end{pmatrix} \Psi_{n}-\frac{\Delta_{n}^2(t)}{2}. \label{GN discrete HS}
\end{equation}
As in the continuous case, we may recover the original Lagrangian after integrating out the HS field. We work in the large-$N$/mean field limit, and thus assume that the HS parameter is time-independent. The Lagrangian is quadratic in the fermionic field, so we may integrate out the fermions to get the effective action for $\Delta$:
\begin{equation} \label{effective action discrete}
    S_{\text{eff}}(\Delta) = S_1(\Delta)-\log\int \mathcal{D}\Psi \mathcal{D}\Psi^{\dagger} \exp\left[-S_2\left(\Psi, \Delta\right)\right], 
\end{equation}
where
\begin{align}
    S_1(\Delta) &= -\frac{T}{2} \sum_{n} \Delta_n^2 \label{S1 eqn}\\
    S_2(\Psi, \Delta) &=  \sum_{n}\int_{0}^{T} dt\, \Psi_{n}^{\dagger} \begin{pmatrix} i\frac{\partial}{\partial t} & \partial_{n}+g\Delta_{n} \\ \partial_{n}^{\dagger}+g\Delta_{n} & i\frac{\partial}{\partial t} \end{pmatrix} \Psi_{n}. \label{S2 eqn}
\end{align}
Note that the integration measure in the path integral is defined as $\mathcal{D}\psi = \prod_n d\psi_n$. We will show that the effective action can be written as a function of the scattering data according to the following equation:
\begin{equation} \label{Effective action as scattering data}
    \frac{S_{\text{eff}}}{T} = \frac{1}{4 \pi g^2} \int_{0}^{2\pi}\log\left(1-|r(e^{i\phi})|^2\right)\cos{(\phi)}\,d\phi - \frac{1}{g^2}\sum_{j=1}^M\left(z_j + z_j^{-1}\right) + 2N \int_{0}^{2\pi} \frac{d\phi}{2\pi}\frac{\sin(\phi)}{\sqrt{u-2\cos{\phi}}}\delta_{t}(\phi),
\end{equation}
where $r$ is the reflection coefficient of the scattering problem associated with the equations of motion, $z_j$ is related to the energy of the bound states, $M$ is the number of bound states and $\delta_{t}(\phi)$ is the phase of the transmission coefficient. In the rest of this subsection, we will derive the above equation and analyze its saddle point solutions. First note that from the action in Eqn. \eqref{GN discrete HS}, the equations of motion for the fermionic fields $\psi_{n}^{1}$ and $\psi_{n}^{(2)}$ are
\begin{align} \label{psi_1 EOM}
    i\frac{\partial \psi_{n}^{(1)}}{\partial t} & = -g\Delta_n\psi_{n}^{(2)} - \partial_n \psi_{n}^{(2)} \triangleq \left(-A \psi^{(2)}\right)_n, \\ \label{psi_2 EOM}
    i\frac{\partial \psi_{n}^{(2)}}{\partial t} & = -g\Delta_n\psi_{n}^{(1)} - \partial_n^{\dagger} \psi_{n}^{(1)} \triangleq \left(-A^\dagger \psi^{(1)}\right)_n,
\end{align}
where $A$ and $A^\dagger$ are matrices whose actions on the fields are encoded in the relations above. Note that there are a total of $N$ pairs of equations of the above form, one for each flavor of fermion. If $\Delta_n$ is time-independent, taking a time derivative of Eqn. \eqref{psi_1 EOM} and \eqref{psi_2 EOM} gives the equations of motion $\ddot{\psi}^{(1)} = - A A^\dagger \psi^{(1)}$ and $\ddot{\psi}^{(2)} = - A^\dagger A \psi^{(1)}$. These equations of motion have a supersymmetric structure, which will be further used to construct soliton solutions in the next section.  We may map these equations onto a discrete time-independent Schr{\"o}dinger equation of the form
\begin{equation} \label{Schrod: Discrete}
    -(t_{n+1}^{(i)}\psi_{n+1}^{(i)}+ t_{n}^{(i)}\psi_{n-1}^{(i)}) + u_n^{(i)}\psi_n^{(i)} = \omega^2 \psi_n^{(i)},
\end{equation}
where $i=1,2$ indicates the index of fermions.  Due to the supersymmetric structure, the hoppings have the form
\begin{equation} \label{hopping}
    t_{n}^{(1)} = g \tau_{n} \Delta_{n-1}, \hspace{0.5cm} t_{n}^{(2)} = g \tau_{n} \Delta_{n},
\end{equation}
and the onsite potentials have the form 
\begin{align} \label{onsite}
    u_{n}^{(1)} = \tau_{n}^2+g^2\Delta_n^2, \hspace{0.5cm}
    u_{n}^{(2)} = \tau_{n+1}^2 +g^2\Delta_n^2.
\end{align}
From now on we will perform the calculation for $i = 1$ and drop the superscript $(i)$ in Eqn. \eqref{Schrod: Discrete}.  This is because the associated scattering problems for $i = 1$ and $i = 2$ have nearly identical scattering data and it does not matter which data we choose to write the action in terms of.

 For the discrete Gross-Neveu model we will need only the first identity with $p=1$:
\begin{equation} \label{p=1 identity}
     \sum_{n=-\infty}^{\infty} u_{n} = -\frac{1}{2\pi} \int_{0}^{2\pi}\log{\left[1-|r(e^{i\phi})|^2\right]}\cos{(\phi)}\,d\phi +\sum_{j=1}^{M}\left(z_j-\frac{1}{z_j}\right).
\end{equation}
For the Gross-Neveu model with the potential $u_{n} = \tau_{n}^2+g^2\Delta_n^2$, we find that $\sum_{n=-\infty}^{\infty} u_{n} = g^2 \sum_{n=-\infty}^{\infty}\Delta_{n}^2 + \Upsilon$, where $\Upsilon = \sum_n \tau_n^2$ is a fixed constant that we will henceforth ignore (more precisely, it can be compensated by an appropriate renormalization of the vacuum energy in the original field theory).  Thus, we can write the first term of the effective action $S_1$ given by Eqn. \eqref{effective action discrete} as a function of the scattering data:
\begin{equation} \label{S2 scattering data}
    S_1 = -\frac{T}{2} \sum_{n}\Delta_n^2 = -\frac{T}{2g^2}\left[-\frac{1}{2\pi} \int_{0}^{2\pi}\log{\left[1-|r(e^{i\phi})|^2\right]}\cos{(\phi)}\,d\phi  + \sum_{j=1}^M\left(z_j-\frac{1}{z_j}\right)\right].
\end{equation}
Now we will express the remaining part of the action as a function of the scattering data. We first need to evaluate the following path integral which is quadratic in the fermions:
\begin{equation} \label{path integral}
   \mathcal{I}_{N}(\Delta) = {\int \mathcal{D}\Psi \mathcal{D}\Psi^{\dagger} \exp\left[-S_2(\Psi, \Delta)\right]},
\end{equation}
where again, $N$ is the number of flavors of fermions and $S_2$ is given by Eqn. \eqref{S2 eqn}. The fermionic fields $\psi^{(1,2)}$ satisfy the anti-periodic boundary conditions in time: $\psi^{(1,2)}(t+T) = -\psi^{(1,2)}(t)$. Following Dashen \textit{et al.} \cite{Neveu_PRD_1975}, the path integral can be written down as
\begin{equation}\label{value of path integral}
    \mathcal{I}_{N}(\Delta) = \left[\mathcal{I}_{1}(\Delta)\right]^{N} =  \mathcal{I}_{1}(0)\prod_{m,k}\frac{\epsilon_{m,k}(\Delta)}{\epsilon_{m,k}(0)},
\end{equation}
where $\epsilon_{m,k}$ are the eigenvalues of the following matrix relation:
\begin{equation} \label{matrix eqn for path integral}
    \begin{pmatrix} i\frac{\partial}{\partial t} & \partial_{n}+g\Delta_{n} \\ \partial_{n}^{\dagger}+g\Delta_{n} & i\frac{\partial}{\partial t} \end{pmatrix} \begin{pmatrix} \xi_{m,k}^{(1)} \\ \xi_{m,k}^{(2)} \end{pmatrix} = \epsilon_{m,k}\begin{pmatrix} \xi_{m,k}^{(1)} \\ \xi_{m,k}^{(2)} \end{pmatrix}.
\end{equation}
and $\mathcal{I}_{1}(0) = \prod_{m,k} \epsilon_{m,k}(0)$ is independent of the onsite potential and hopping.  Suppose $\phi_{k}^{(1,2)}$ satisfies Eqn. \eqref{matrix eqn for path integral} where the right hand side is 0. For a time-independent $\Delta_n$, we may choose $\phi_{k}^{(1,2)}$ such that it satisfies $\phi_{k}^{(1,2)}(t+T) = e^{-i\alpha_{k}(\Delta)}\phi_{k}^{(1,2)}(t)$ where $\alpha_{k}(\Delta) = \omega_{k}(\Delta) T$ and $\omega_k^2$ is the energy. In general, for $\Delta_n(t)$ time periodic with period $T$, $\alpha_k$ label the Floquet indices; however, when $\Delta_n$ is time independent, the Floquet indices coincide with the energy.  One can show that the function $\xi_{m,k}^{(1,2)}(t)=\exp\left(i(2m+1)\pi t/T+i \alpha_{k} t/T\right) \phi_{k}^{1,2}(t)$ satisfies the anti-periodic boundary condition and is an eigenfunction of Eqn. \eqref{matrix eqn for path integral} with eigenvalue:
\begin{equation} \label{eigenvalue}
    \epsilon_{m,k} = -\frac{(2m+1)\pi}{T} - \frac{\alpha_k}{T}
\end{equation}
We substitute Eqn. \eqref{eigenvalue} into Eqn. \eqref{value of path integral} and simplify it using the standard infinite product identity for cosine to get:
\begin{align}
     \mathcal{I}_{1}(\Delta) &= \mathcal{I}_{1}(0) \prod_{k}(e^{i\alpha_{k}/2}+e^{-i\alpha_{k}/2}) \\
     &= \mathcal{I}_{1}(0)\exp\left(i\sum_{k}|\alpha_{k}|\right)\prod_{\alpha_k > 0}(1+e^{-i\alpha_{k}})^2 \\
     &= \mathcal{I}_{1}(0)\exp\left(i\sum_{k}|\alpha_{k}|\right)\sum_{\{n_i\}} e^{-i\sum_{\alpha_i>0} n_i\alpha_{i}} \prod_{i} q_{n_i}
\end{align}
where $0 \leq n_i \leq 2N$ is the population of the $i$th energy mode, $q_{n_i} = \frac{2N!}{n_i!(2N-n_i)!}$, and in the second line we have used the fact that the energies come in pairs with opposite sign to restrict the summation to positive energy modes. In general, we may assume that the set of Floquet indices are comprised of a set of isolated, discrete states and a continuum of scattering states. The $N$ fermions and $N$ anti-fermions may occupy these states, and we will assume that the saddle point corresponds to $n_0$ fermions occupying a single discrete state $\alpha_0$. Thus, we may replace the summation over $n_i$ with a constant $n_0$ and write the action as:
\begin{equation}
    S_{\text{eff}} = S_1 + N\sum_{k} |\alpha_k| - n_0 \alpha_0 + \text{const.}
\end{equation}
We have already written the second term $S_1$ as a function of the scattering data. Now we want to do the same for the first term $\sum_{k} |\alpha_k|$. We have noticed that for time-independent potentials $\alpha_k = \omega_k T$ where $\omega_k^2=u-2\cos{k}$ is the energy as a function of quasi-momentum. In Appendix \ref{Appendix B} we show that for a system with $N$ sites the quasi-momentum $k$ is quantized as $k_{n}^{\pm} = \frac{2\pi n-\lambda^{\pm}}{N}$ where $\lambda^{\pm}$ are the eigenvalues of the scattering matrix.  Then, one may show, as derived in the appendix:
\begin{equation} \label{appendixb}
    \sum_{k}|\alpha_k| = 2T \int_{0}^{2\pi} \frac{d\phi}{2\pi}\frac{\sin(\phi)}{\sqrt{u-2\cos{\phi}}}\delta_{t}(\phi) + \omega_0 T + \Lambda,
\end{equation}
where $\delta_{t}(\phi)$ is the phase of the transmission coefficient and $\Lambda$ is a divergent constant and can be tamed by a suitable vacuum energy shift. We have also accounted for the presence of the single bound state $\alpha_0 = \omega_0 T$.  The transmission coefficient is the inverse of $a(z)$, and so the phase of the the transmission coefficient is the imaginary part of $\log{a^{-1}(z)}$. Using the Jensen formula like before, we can write an equation for $\log{a(z)}$ similar to Eqn. \eqref{ln(a) and scattering data: Discrete}:
\begin{equation} \label{JensenIm}
    \delta_{t}(\theta) = -\sum_{j=1}^{N}\Im\left(\log {\frac{z-z_j}{1-z_jz}}\right) + \frac{1}{4\pi}\int_{0}^{2\pi}\Im\left( \frac{e^{i\phi}+z}{e^{i\phi}-z}\right)\log\left[1-|r(\phi)|^2\right]\,d\phi, 
\end{equation}
where $z=e^{i\theta}$ and $z_i$ are the zeros of $a(z)$.  Note that $\Im(z)$ denotes the imaginary part of $z$. Combining all the results together we get an equation for effective action as a function of the scattering data:
\begin{multline} \label{Action in terms of scattering data}
    \frac{S_{\text{eff}}}{T} = \frac{1}{4\pi g^2} \int_{0}^{2\pi}\log\left(1-|r(e^{i\phi})|^2\right)\cos{(\phi)}d\phi  -\frac{1}{2g^2} \sum_{j=1}^N\left(z_j-\frac{1}{z_j}\right) \\+ 2N \int_{0}^{2\pi} \frac{d\phi}{2\pi}\frac{\sin(\phi)}{\sqrt{u-2\cos{\phi}}}\delta_{t}(\phi) + (N - n_0)\omega_0.
\end{multline}
Upon varying the action with respect to scattering data, we obtain the saddle point solutions for the HS parameter, which correspond to solitonic textures.  Setting $\delta S_{\text{eff}}/\delta r = 0$ gives the solution $r(\phi) = 0$, which implies that the effective onsite potential $u_n = \tau_{n}^2+g^2\Delta_n^2$ must admit reflectionless scattering solutions. However there is an additional subtlety in the discrete case that does not appear in the continuous case. We note that in Eqn. \eqref{Action in terms of scattering data} there is a $\cos \phi$ multiplying $\log\left(1-|r(e^{i\phi})|^2\right)$ in the integrad which was absent in the continuous case. Consequently, one can construct a potential which, in principle, has a smaller contribution to the action than a reflectionless potential.  However, there are no physical potentials that we are aware of that have such scattering properties, so we deem that such kinds of potentials are likely unphysical.  

One obvious example of a reflectionless potential is a constant; however, it is well known that there are many nontrivial candidates which are also reflectionless and give rise to kink solitons.  To construct such nontrivial solutions, we will utilize a version of supersymmetric quantum mechanics for tight binding models, which will be discussed in the next section.  We will show that we can construct a hierarchy of reflectionless soliton solutions by iterating the supersymmetric transformation starting from a constant onsite potential.  

\begin{figure}
    \begin{tabular}{cc}
  \includegraphics[scale=0.685]{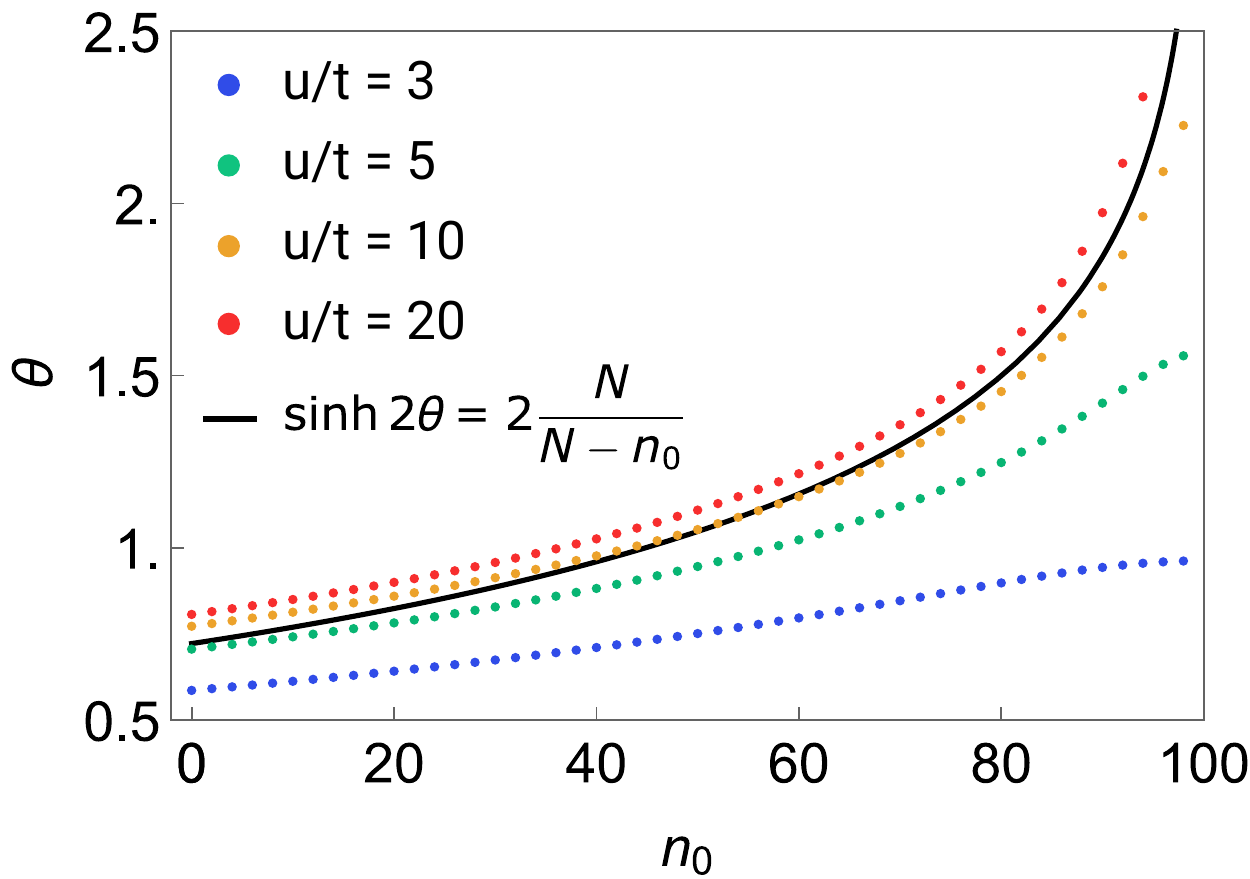} &   \includegraphics[scale=0.65]{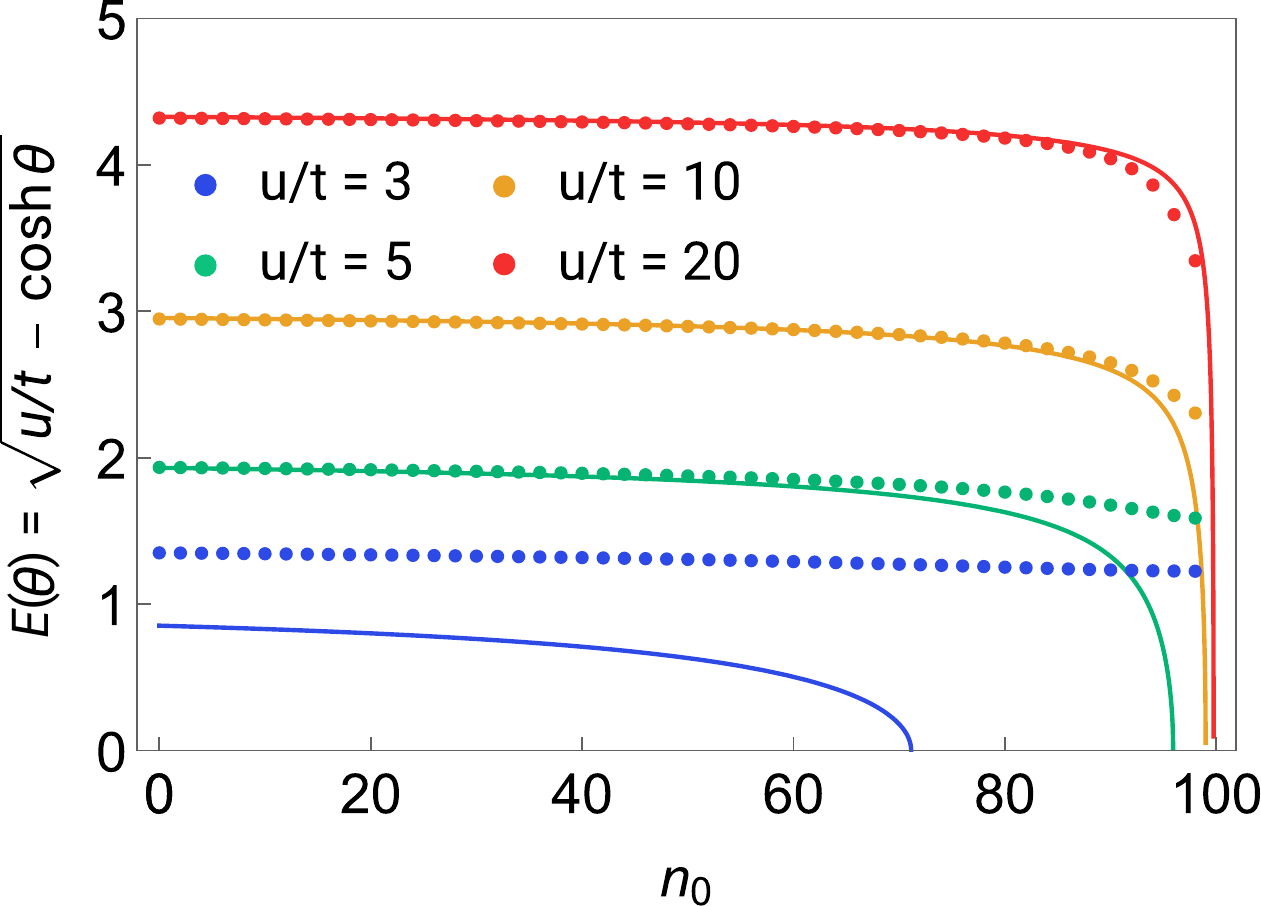} \\
    \end{tabular}
    \caption{In the left panel, a plot of the values of $\vartheta$ as a function of the filling $n_0$.  Here, $g = 1$ and we have chosen $N = 100$ and varied $u$.  Also plotted in black is the approximation $\sinh 2\vartheta \approx \frac{2 N}{(N - n_0)}$, which is a valid in a regime where $u$ is large compared to $\cosh \theta$ but small compared to $N^2$.  We even see good accuracy of the approximation beyond the regime of validity.  In the right panel, a plot of the bound state energy as a function of the filling $n_0$.  The solid lines are the corresponding bound state energies obtained from the approximation.}
    \label{fig:boundspec}
\end{figure}

Furthermore, we must also vary the action with respect to the bound state energy; this results in the mass/energy spectrum for semiclassical bound states in the Gross-Neveu/lattice superconductor model.  There are two separate cases that we must consider, both of which give rise to physically distinct soliton solutions.  
\begin{itemize}
    \item In the first case, we assume that the HS parameter and the hopping converge to the constant values $\Delta_\infty$ and $\tau_\infty$ at both $\pm\infty$; the value of the potential and hopping is related to these asymptotic values via $u = g^2\Delta_\infty^2 + \tau_\infty^2$ and $t = g \Delta_\infty \tau_\infty$, the latter of which has been set to one.  For simplicity, we follow same assumption as in Dashen \textit{et al.} \cite{Neveu_PRD_1975} that there is a single, isolated bound state occupied by $n_0$ fermions (the general case in which we allow for multiple bound states is a straightforward extension of the analysis presented). We calculate the energy of that bound state at the saddle point, which is given by setting $\delta S_{\text{eff}}/\delta z_1 = 0$ with the reflection coefficient $r(\phi) = 0$. From Eqn. \eqref{Action in terms of scattering data} we get 
    %\textcolor{red}{CHECK SIGNS + RENORMALIZATION?}
\begin{align}
    \frac{1}{T}\frac{\delta S_{\text{eff}}}{\delta z_1} = -\frac{1}{2g^2}\left(1+\frac{1}{z_1^2}\right)-4N\int_{0}^{2\pi} \frac{d\phi}{2\pi}\frac{\sin^2\phi}{\sqrt{u - 2\cos\phi}}&\frac{1}{1+z_1^2-2z_1\cos{\phi}} \nonumber \\  &+ (N-n_0)\frac{1-z_1^2}{2 z_1^2\sqrt{u - (z_1 + 1/z_1)}}.
\end{align}
We may perform a convenient change of variables $2 \cosh \vartheta = 1/z_1+ z_1$ and $2 \sinh \vartheta = 1/z_1 - z_1$. Then, the condition $\delta S_{\text{eff}}/\delta z_1 = 0$ becomes
\begin{equation}
     -\frac{1}{g^2} \cosh \vartheta -2N\int_{0}^{2\pi}\frac{d\phi}{2\pi}\frac{\sin^2\phi}{\sqrt{u - 2\cos\phi}}\frac{1}{\cosh \vartheta-\cos{\phi}} + (N-n_0)\frac{\sinh \vartheta}{\sqrt{u - 2 \cosh \vartheta}} = 0.
\end{equation}
The integral has a closed form in terms of elliptic functions, but the subsequent implicit equation for $\cosh \vartheta$ is difficult to analyze analytically.  We have evaluated the integral numerically and found that for a wide range of parameters there is only one solution where $\cosh \vartheta > 1$.  To make progress analytically, we proceed by assuming that $\cosh \vartheta \gg 1$, which physically assumes that the bound state is strongly bound; then, the equation above can be simplified to
\begin{equation}
     -\frac{1}{g^2} \cosh \vartheta - \frac{N}{\cosh \vartheta} I(u) + (N-n_0)\frac{\sinh \vartheta}{\sqrt{u - 2 \cosh \vartheta}} = 0,
\end{equation}
where
\begin{equation}
    I(u) = 2 \int_{0}^{2\pi}\frac{d\phi}{2\pi}\frac{\sin^2\phi}{\sqrt{u - 2\cos\phi}}.
\end{equation}
This may be reduced to a fifth order polynomial equation in $\cosh \vartheta$ which is still not analytically tractable.  In the limit where $u \gg 2 \cosh \vartheta$ and $\sqrt{u} \ll g^2 N$, we may neglect the first term and utilize the simplification $I(u) = 1/\sqrt{u}$.  Then, we obtain that $\sinh 2\vartheta \approx \frac{2 N}{(N - n_0)}$, which is an approximately valid solution for moderate values of the filling $n_0/N$.  We plot both the approximation and exact values of $\theta$ in Figure \ref{fig:boundspec}, and it is clear that the approximation holds even beyond the regime of validity described.

The corresponding reflectionless potentials with these bound state energies are nontopological solitons which are kink-antikink solitons.  These solitons are strongly dependent on the number of trapped fermions; intuitively, this is because repulsion due to the $n_0$ fermions trapped in the kink and antikink will cause them to separate; the calculation above self consistently determines the stable value for the separation distance, which is also related to the bound state energy.  

\item In the second case, we assume that the HS parameter and the hopping converge to different values $\Delta_{\pm\infty}$ and $\tau_{\pm\infty}$ at $\pm\infty$; we find that these solitons are topological, and are associated with the Callan-Coleman-Gross-Zee (CCGZ) kink solitons first discovered famously in the $\varphi^4$-field theory.  If we assume that $n_0$ of the fermions are trapped in a single bound state, we arrive at a slightly different result assuming kink boundary conditions.  First, we note that the scattering problems for both components of the fermion differ by a bound state.  This can be seen by computing 
\begin{equation}
    \sum_n (u_n^{(2)} - u_n^{(1)}) = \sum_n (\tau_{n+1}^2 - \tau_n^2) = \tau_{\infty}^2 - \tau_{-\infty}^2,
\end{equation}
and therefore the right hand side of the trace identity in Eqn. \eqref{p=1 identity} will differ between the two scattering problems.  We will show in the next section that the reflection coefficient of the scattering problems are the same, and so the two scattering problems must differ in the number of bound states.  For there to be a single bound state, the scattering problem for the first component of the fermion must have no bound states and must be reflectionless.  The action reduces drastically to $S_{\text{eff}} = (N-n_0) \omega_0$, which is minimized for $\omega_0 = 0$ regardless of the value of $n_0$.  The intuition for this is rather simple.  The theory possesses a topological zero-mode, the Jackiw-Rebbi mode, which is insensitive to the number of occupied fermions; therefore, the nature of this kink soliton should not depend on the filling.

We also know that the bound state energies are given by the explicit formula 
\begin{equation}
    \omega_0 = \sqrt{u - z_1 - \frac{1}{z_1}}.
\end{equation}
Using the formulae $u = g^2 \Delta_\infty^2 + \tau_\infty^2 = g^2 \Delta_{-\infty}^2 + \tau_{-\infty}^2 $ and $t = 1 = g \Delta_\infty \tau_\infty = g \Delta_{-\infty} \tau_{-\infty}$, we find that $z_1 = \tau_\infty^2$.  Substitution into $\omega_0$ also gives $\omega_0 = 0$, consistent with the saddle point solution obtained directly from minimizing the action.
\end{itemize}

\subsection{Construction of field theory hierarchy}
The lattice Gross-Neveu model presented in the previous section can be considered to be the first such model in a hierarchy of field theories that admit soliton solutions.  First note that we rather naturally divided the action for the lattice GN model into two pieces given by Eqns. \eqref{S1 eqn} and \eqref{S2 eqn}. The first part of the action was rewritten using the first trace identity, while no trace identities were required for the second part of the action. To generalize, we write the first part of the action as a trace identity of our choice; the form of such an action is
\begin{equation}
    S = -\sum_{n}\int_{0}^{T} dt\, F(\Delta_n(t), \tau_n) + \sum_{n}\int_{0}^{T} dt\, \Psi_{n}^{\dagger} \begin{pmatrix} i\frac{\partial}{\partial t} & \partial_{n}+g\Delta_{n}(t) \\ \partial_{n}^{\dagger}+g\Delta_{n}(t) & i\frac{\partial}{\partial t} \end{pmatrix} \Psi_{n},
\end{equation}
where $F(\Delta_n,\tau_n)$ is constructed from the trace identities, and can be written as a linear combination of traces of powers of the Hamiltonian. If we utilize the first trace identity and the potential given by Eqn. \eqref{onsite} to write $F(\Delta_n, \tau_n)=u_n=g^2\Delta_n^2$ then we get back the lattice GN model. In general if we use any trace identity, the action will have a saddle point corresponding to a reflectionless potential because the trace identities have a $\log(1-|r|^2)$ dependence on the reflection coefficient $r$; thus a solution to $\delta S_{\text{eff}}/\delta r = 0$ will always be $r=0$. To exemplify this process, we will construct an action using the second trace identity. In this case we choose $F(\Delta_n,\tau_n)$ to be
\begin{equation} \label{F for trace iden 2}
    F(\Delta_n,\tau_n) = u_n^2+2t_n^2 = \tau_n^4 + g^4\Delta_n^4 + 2 g^2 \Delta_n^2\left(\tau_n^2 + \tau_{n+1}^2\right),
\end{equation}
where we have used Eqn. \eqref{onsite} and \eqref{hopping} to write down the form of the onsite potential and hopping. Substituting Eqn. \eqref{F for trace iden 2} into the action we get a new field theory with a background field $\tau$ and the Hubbard-Stratonovich field $\Delta$. First we assume that the HS field is time-independent as before.  Then, we may integrate out the HS field to construct a local theory in the fermions.  Upon performing the integration, the action becomes
\begin{equation}
    S = \sum_{n}\int_{0}^{T} dt\, \Psi_{n}^{\dagger}(t) \begin{pmatrix} i\frac{\partial}{\partial t} & \partial_{n} \\ \partial_{n}^{\dagger} & i\frac{\partial}{\partial t} \end{pmatrix} \Psi_{n}(t) + \log(Z),
\end{equation}
where $Z$ is obtained by taking the path integral of the remaining part of the action with respect to the HS field:
\begin{equation}
    Z = \prod_{n} \int_{-\infty}^{\infty} d\Delta_n \exp{\left[\Psi_{n}^{\dagger} \begin{pmatrix} 0 & g\Delta_n \\ g\Delta_n & 0 \end{pmatrix} \Psi_{n}- g^4\Delta_n^4 - 2 g^2 \Delta_n^2\left(\tau_n^2 + \tau_{n+1}^2\right)\right]},
\end{equation}
where we have ignored the constant term coming from $\tau_n^4$.  The functional $Z$ can be further simplified to the following form
\begin{equation}
    Z \propto \exp{\left(\sum_n \frac{\zeta\left(\tau_n^2+\tau_{n+1}^2\right)}{2g^2}\left(\overline{\Psi}_n\Psi_{n}\right)^2\right)}.
\end{equation}
Therefore, the full action is a Gross-Neveu model where the interaction has site-dependence and can be treated as a background field.  The function $\zeta$ is given by
\begin{align}
    \zeta\left(\tau_n^2+\tau_{n+1}^2\right) &= \frac{\int_{-\infty}^{\infty} dx\,  x^2e^{-x^4 - \left(\tau_n^2+\tau_{n+1}^2\right)x^2}}{\int_{-\infty}^{\infty} dx\, e^{-x^4 - \left(\tau_n^2+\tau_{n+1}^2\right)x^2}},
\end{align}
which may be expressible in terms of special functions.  In principle, one can treat $\tau$ as a scalar field in the path integral formulation; in this case, the full theory is obtained by integrating over $\tau_n$.  Unfortunately, since the dependence of the effective interaction on the $\tau_n$'s does not decouple, the resulting theory will look quite non-local.  This is analogous to what happens in the continuum limit, except there the non-locality of the theory arises when integrating out the HS parameter.

If one continues to construct more such field theories using higher order trace identities, the corresponding actions will take the generic form
\begin{equation}
    S = \int_{0}^{T} dt\, \left[\sum_{n}\, \Psi_{n}^{\dagger}(t) \begin{pmatrix} i\frac{\partial}{\partial t} & \partial_{n} \\ \partial_{n}^{\dagger} & i\frac{\partial}{\partial t} \end{pmatrix} \Psi_{n}(t) + \zeta_k\left(\tau_n, \tau_{n+1},\cdots,\tau_{n+k-1}\right)\left(\overline{\Psi}_n(t)\Psi_{n}(t)\right)^2\right],
\end{equation}
where $k$ indexes the action constructed using the $k$th trace identity.  It is a rather elegant feature that nontrivial field theories with soliton solutions may be reverse engineered from a series of trace identities; however, for each of these models the mass spectrum (or equivalently the bound state/soliton spectrum) will vary depending on which trace identity is used.  Such a calculation is a simple modification of the procedure developed in the previous subsection.

\section{Exact construction of saddle point solutions}
In the previous sections, we justified that a saddle point of the action for the lattice Gross-Neveu model corresponds to reflectionless potentials with a particular spectrum of bound states.  To obtain the original potential corresponding to a particular set of scattering data, the inverse scattering method is applied, which involves solving the Gelfand-Levitan equation.  Since this formulation can be rather bulky, in this section we introduce an alternate formalism based on a generalization of supersymmetric quantum mechanics to tight binding models, and use this method to construct the kink soliton and kink-antikink solutions described in the previous section, without resorting to using the Gelfand-Levitan equation.

\subsection{Continuous supersymmetry} \label{Cont SUSY}
Here, we briefly review the well-established field of supersymmetric (SUSY) quantum mechanics \cite{COOPER1995267}.  We start with the Schr\"odinger operator factorized as $\mathcal{H} = -\partial_x^2 + V(x) = A^\dagger A$, where $A^\dagger = -\partial_x + W(x)$ and $A = \partial_x + W(x)$.  By convention, $W(x)$ is known as the superpotential, and is related to the original potential via the Riccati equation $W^2(x) - W'(x) = V(x)$.  We then construct a new Hamiltonian $\widetilde{\mathcal{H}} = A A^\dagger$, which is also a valid Schr\"odinger operator with a new potential $\widetilde{V}(x) = W^2(x) + W'(x)$.  The Hamiltonian $\widetilde{\mathcal{H}}$ has the same spectrum as the Hamiltonian $\mathcal{H}$, apart from the removal of a zero energy eigenstate, if any, which solves $A \psi = 0$.  Furthermore, if $\psi(x)$ is an eigenstate of $\mathcal{H}$ with energy $E$, then $A \psi(x)$ is an eigenstate of $\widetilde{\mathcal{H}}$ with the same energy.

The canonical example of continuous soliton solutions can be obtain by repeated application of the SUSY method to a constant potential.  In particular, if we consider the Schr\"odinger equation
\begin{equation}
    \left(-\frac{\partial^2}{\partial x^2} - N(N+1) \, \text{sech}^2 x\right) \psi(x) = E \psi(x),
\end{equation}
we may find that the system at $N$ can be related to the system at $N+1$ via a supersymmetry transformation with $A = -\partial_x - N \tanh x$; the case $N = 0$ corresponds to a constant potential.  One can show that the number of bound states exactly equals $N$ and that their energies are given by $E_n = -(n - N)^2$.  The scattering states where $E > 0$ are reflectionless. This is an example of an $N$-soliton solution with the property that the number of solitons coincides with the number of bound states.

\subsection{Discrete supersymmetry: formalism}

We would like to extend the formalism of continuous supersymmetry to tight binding models.  The simplest structure of a Hamiltonian which is amenable to supersymmetry is a tridiagonal matrix, equivalent to the one from Eqn. \eqref{discreteschroham}, with $t_n$ the hopping amplitudes and $u_n$ the onsite potentials.  We factor $\mathcal{H} = A^\dagger A$, and select an ansatz for $A$ to be
\begin{equation}
    A^\dagger = \sum_{n = -\infty}^\infty \tau_{n+1} |n\rangle \langle n+1| + \Delta_n |n\rangle \langle n|,
\end{equation}
which is a discrete version of $A^\dagger = -\partial_x + W(x)$; we will call $\tau$ the superhopping and $\Delta$ the superpotential.  Note that the matching notation with the previous section is no coincidence: indeed, the matrix $A^\dagger$ is equal to $\partial^\dagger_n + \Delta_n$ (we will set $g = 1$ for convenience).  The equations of motion of the Gross-Neveu model decouple into independent equations of motion for the effective Hamiltonians $\mathcal{H} = A^\dagger A$ and $\widetilde{\mathcal{H}} = A A^\dagger$, both of which are tridiagonal and have nearly identical spectra because they are superpartners.  Crucially, knowing $\mathcal{H}$, one may extract the superhopping and superpotential and use it to compute $\widetilde{\mathcal{H}}$.

We can iterate this process to create a family of nearly isospectral Hamiltonians, which gives us a hierarchy of saddle point solutions to the Gross-Neveu model.  For example, at the $k$th iteration, we construct the Hamiltonian $\mathcal{H}_k = A_k A_k^\dagger$ and solve the matching condition $A_k A_k^\dagger = A_{k+1}^\dagger A_{k+1}$ to obtain a new set of superpotentials and superhoppings.  The Hamiltonian at the $(k+1)$st iteration can then be constructed via $\mathcal{H}_{k+1} = A_{k+1} A_{k+1}^\dagger$.  In this hierarchy, any pair of consecutive Hamiltonians can be considered to be a valid saddle point solution for the Gross-Neveu model.  The matching condition can be obtained from Eqns. \eqref{hopping} and \eqref{onsite}:
\begin{align}
    \tau_{n}^{(k-1)} \Delta_{n-1}^{(k-1)} &= \tau_{n}^{(k)} \Delta_{n}^{(k)} \label{SUSY hierarchy},\\
    {\tau_{n}^{(k-1)}}^2+ {\Delta_{n}^{(k-1)}}^2 &= {\tau_{n+1}^{(k)}}^2 +  {\Delta_{n}^{(k)}}^2 \label{SUSY hierarchy2}.
\end{align}
Note that the subscripts denote the location on the lattice while the superscripts indicate the level of the hierarchy.  The Hamiltonians $\mathcal{H}_k$ and $\mathcal{H}_{k+1}$ have the same spectrum and differ by a possible zero-energy state that appears only in the limit of an infinitely long chain with open boundary conditions; this zero-energy state is analogous to the one in continuous supersymmetry.  Using this method, one can create a Hamiltonian with arbitrarily placed energy levels for the bound states: one only needs to introduce the constant shift $\mathcal{H}_{k} \to \mathcal{H}_{k} + \delta_{k}$ after each iteration.  The corresponding bound state energy levels after the $k$th iteration will be $E_{m} = \sum_{i=1}^m \delta_i$ for $m \leq k$.  As a result, this iterative method can be used to obtain saddle point solutions given a sequence of bound state energies; we need only prove that the corresponding potentials are reflectionless.  To see this, first note the eigenstates of the Hamiltonian after $k$ levels are given by
\begin{equation}
    |\psi\rangle_k \propto A_k A_{k-1} \ldots A_1 |\psi\rangle_1,
\end{equation}
so that due to the local properties of the matrix $A$, bound states remain bound and scattering states remain scattering.  Next, we consider a scattering process that occurs in the base Hamiltonian $\mathcal{H}_1$; assuming the potential and hopping are asymptotically constants $u$ and $t$ and the incident scattering particles come from the left, we may write the wavefunction using $z = e^{ik}$ as
\begin{equation}
    \psi_n(z) = 
  \begin{cases}
    z^n + r(z) z^{-n}, & \text{for } n \to -\infty 
    \\
    t(z) z^n, & \text{for } n \to \infty
  \end{cases}
\end{equation}
where $t(z)$ and $r(z)$ are the transmission and reflection coefficients.  Upon supersymmetry, the new scattering eigenstates in $\mathcal{H}_2$ are $\overline{\psi}_n(z) = \Delta_n \psi_n(z) + \tau_{n+1} \psi_{n+1}(z)$, which can be written as
\begin{equation}
    \overline{\psi}_n(z) = 
  \begin{cases}
    z^n + r(z) \frac{z \Delta_{-} + \tau_{-}}{z(\Delta_{-} + \tau_{-} z)} z^{-n}, & \text{for } n \to -\infty 
    \\[10pt]
    t(z) \frac{\Delta_{+} + \tau_{+} z}{\Delta_{-} + \tau_{-} z} z^n, & \text{for } n \to \infty
  \end{cases}
\end{equation}
where the subscripts $+$ and $-$ denote the asymptotic values of the superpotential and superhopping at positive and negative infinity respectively.  which allows us to identify the modified reflection and transmission coefficients $\overline{r}(z)$ and $\overline{t}(z)$.  In particular, we find that
\begin{equation}
    |\overline{r}|^2 = \frac{|r|^2}{|z|^2} \frac{(\tau_{-} + \Delta_{-}\cos \phi)^2 + \Delta_{-}^2 \sin^2 \phi}{(\Delta_{-} + \tau_{-}\cos \phi)^2 + \tau_{-}^2 \sin^2 \phi} = |r|^2,
\end{equation}
and
\begin{equation}
    |\overline{t}|^2 = |t|^2 \frac{(\Delta_{+} + \tau_{+}\cos \phi)^2 + \tau_{+}^2 \sin^2 \phi}{(\Delta_{-} + \tau_{-}\cos \phi)^2 + \tau_{-}^2 \sin^2 \phi}
    = |t|^2 \frac{{\Delta_{+}}^2 + {\tau_{+}}^2 + 2 \Delta_{+} \tau_{+}\cos \phi}{{\Delta_{-}}^2 + {\tau_{-}}^2 + 2 \Delta_{-} \tau_{-}\cos \phi} = |t|^2,
\end{equation}
where we have used the supersymmetry relations ${\Delta_{\pm}}^2 + {\tau_{\pm}}^2 = u$ and $\Delta_{\pm}\tau_{\pm} = t$.  Thus, the superpartner Hamiltonian preserves the magnitude of the reflection and transmission coefficients, and induction proves that this holds for all subsequent Hamiltonians in the hierarchy.  We also note that the poles of $\overline{t}$ are the poles of $t$ along with an additional pole at $z_* = -\Delta_{-}/\tau_{-}$.  This additional pole corresponds to a normalizable bound state at zero energy if $z_*$ lies within the unit disk, or if $\left| \Delta_{-}/\tau_{-} \right| < 1$.

\subsection{Construction of kink solitons}

\begin{figure*}
\begin{tabular}{cc}
  \includegraphics[scale=0.575]{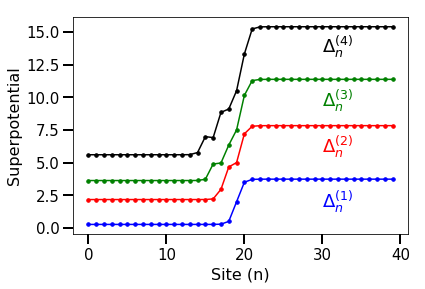} &   \includegraphics[scale=0.575]{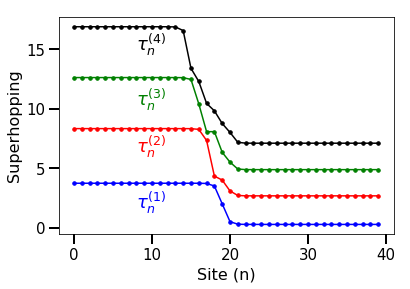} \\
\end{tabular}
\caption{A plot of the superpotential and the superhopping obtained from iterating discrete supersymmetry for four iterations starting with a constant potential.  A constant shift is applied between iterations, and the corresponding solutions share the same spectrum apart from the constant shift and an additional bound state.  These solutions correspond to one such hierarchy of kink solitons.}
\label{fig:topsol}
\end{figure*}

In continuous supersymmetry, a soliton hierarchy is achieved by starting with a reflectionless constant potential and iterating the supersymmetry procedure.  In analogy to the continuous case, let us assume a tight binding Hamiltonian given by
\begin{equation}
    \mathcal{H} = \sum_{n = -\infty}^\infty \left(-t |n\rangle \langle n+1| + \text{h.c.} + u |n\rangle \langle n|\right).
\end{equation}
We iterate the supersymmetric process once.  For this, we will need to solve the equations $t = \tau_i \Delta_{i-1}$ and $u = \tau_i^2 + \Delta_i^2$.  We need to first determine the asymptotic behavior for the superhopping and the superpotential, which implies solving $\tau^2 + \Delta^2 = u$ and $\tau \Delta = t$.  This yields two solutions
\begin{equation} \label{suppm}
    \Delta_{\pm}^2 = \frac{u}{2}\left(1 \pm \sqrt{1 - \frac{4 t^2}{u^2}}\right),
\end{equation}
and $\tau_{\pm} = t/\Delta_{\pm}$.  Thus, our kink solution will transition from one of these solutions at $-\infty$ to the other solution at $+\infty$.  To solve for the intermediate values, we choose the origin as a seed with $\Delta_0 = \Delta$, determine $\Delta_n$ for $n > 0$ and $n < 0$ separately, and self-consistently determine constraints on $\Delta$ for the existence of a solution.  We may solve for the superpotential and find that it takes the form of a continued fraction:
\begin{equation}
   \Delta_{N}^2 = u-\cfrac{t^2}{u -\cfrac{t^2}{u -\cfrac{t^2}{\ddots - \cfrac{t^2}{\Delta^2}}}}
   \hspace{1cm}
   \Delta_{-N}^2 = \cfrac{t^2}{u -\cfrac{t^2}{u -\cfrac{t^2}{\ddots - \Delta^2}}}
\end{equation}
where the number of repetitions in continued fraction is $N$.  We use a well-known result for the finite continuants of a continued fraction.  Given a continued fraction of the form
\begin{equation}
   F(\eta, \gamma) = \eta_0 + \cfrac{\gamma_1}{\eta_1 +\cfrac{\gamma_2}{\eta_2 +\cfrac{\gamma_3}{\eta_3 + \cdots}}},
\end{equation}
the finite continuants can be expressed in the form $F_N(\eta, \gamma) = p_N/q_N$, where $p_n$ satisfies the recurrence
\begin{equation}
    p_n = \eta_n p_{n-1} + \gamma_n p_{n-2}
\end{equation}
with $p_0 = \eta_0$ and $p_{-1} = 1$, and $q_n$ satisfies the same recurrence with $q_0 = 1$ and $q_{-1} = 0$.  In the present case, focusing on the equation for $W_{N}$, we find that $\eta_N = \Delta^2$ and $\gamma_N = -t^2$ as well as $\eta_i = u$ and $\gamma_i = -t^2$ for $i < N$.  We then find the following solution for the recurrence for $p$:
\begin{equation}
    \begin{pmatrix}
  p_N  \\
  p_{N-1}
\end{pmatrix} = \left(\begin{array}{cc}
  \Delta^2 & -t^2 \\
  1 & 0 \\
\end{array}\right) \left(\begin{array}{cc}
  u & -t^2 \\
  1 & 0 \\
\end{array}\right)^{N-1} \begin{pmatrix}
  u  \\
  1
\end{pmatrix},
\end{equation}
and similarly for $q$:
\begin{equation}
    \begin{pmatrix}
  q_N  \\
  q_{N-1}
\end{pmatrix} = \left(\begin{array}{cc}
  \Delta^2 & -t^2 \\
  1 & 0 \\
\end{array}\right) \left(\begin{array}{cc}
  u & -t^2 \\
  1 & 0 \\
\end{array}\right)^{N-1} \begin{pmatrix}
  1  \\
  0
\end{pmatrix}.
\end{equation}
For $\Delta_{-N}$, we find a similar series of expressions, but it can be shown that they are equivalent to an analytic continuation of the above expressions to negative $N$.  Combining both branches, we find the following solution for $\Delta_N^2$:
%\begin{equation}
%     \Delta_N^2 = \Delta_+^2\Delta_-^2 \frac{\Delta_+^{2N-2}(\Delta_+^2-\Delta^2) + \Delta_-^{2N-2}(\Delta^2-\Delta_-^2)}{\Delta_+^{2N}(\Delta_+^2-\Delta^2) + \Delta_-^{2N}(\Delta^2-\Delta_-^2)}.
%\end{equation}
\begin{equation}
     \Delta_N^2 = \frac{\Delta_-^{2N+2}(\Delta_+^2-\Delta^2) + \Delta_+^{2N+2}(\Delta^2-\Delta_-^2)}{\Delta_-^{2N}(\Delta_+^2-\Delta^2) + \Delta_+^{2N}(\Delta^2-\Delta_-^2)}.
\end{equation}
Notice that $\lim_{N \to \infty} \Delta_N^2 = \Delta_+^2$ and $\lim_{N \to -\infty} \Delta_N^2 = \Delta_-^2$ as well as $\Delta_0^2 = \Delta^2$.  Thus, this superpotential behaves like a discretized version of hyperbolic tangent, which is the superpotential in the continuous limit.  However, we do want to enforce that the above equation does not become negative; an accordingly sufficient constraint on $\Delta^2$ is that $\Delta_-^2 < \Delta^2 < \Delta_+^2$.  The superhopping is correspondingly given by $\tau_N = t/\Delta_{N-1}$ and also behaves like a discrete hyperbolic tangent.  Examples of these soliton profiles are plotted numerically in Figure \ref{fig:topsol}.  We may write expression for the new potential $\widetilde{u}_N = \tau_{N+1}^2 + \Delta_N^2$ and the new hopping $\widetilde{t}_N = \tau_{N}\Delta_N$ after one iteration of supersymmetry:
\begin{align}
    \widetilde{u}_N &= t^2 \frac{\Delta_-^{2N}(\Delta_+^2-\Delta^2) + \Delta_+^{2N}(\Delta^2-\Delta_-^2)}{\Delta_-^{2N+2}(\Delta_+^2-\Delta^2) + \Delta_+^{2N_2}(\Delta^2-\Delta_-^2)} + \frac{\Delta_-^{2N+2}(\Delta_+^2-\Delta^2) + \Delta_+^{2N+2}(\Delta^2-\Delta_-^2)}{\Delta_-^{2N}(\Delta_+^2-\Delta^2) + \Delta_+^{2N}(\Delta^2-\Delta_-^2)} \label{superpartnerconst},\\
    \widetilde{t}_N &= t \frac{\left(\Delta_-^{2N}(\Delta_+^2-\Delta^2) + \Delta_+^{2N}(\Delta^2-\Delta_-^2)\right)^2}{\left(\Delta_-^{2N+2}(\Delta_+^2-\Delta^2) + \Delta_+^{2N+2}(\Delta^2-\Delta_-^2)\right)\left(\Delta_-^{2N-2}(\Delta_+^2-\Delta^2) + \Delta_+^{2N-2}(\Delta^2-\Delta_-^2)\right)},
\end{align}
both of which resemble a discretized version of hyperbolic secant.  Furthermore, we may also formulate a condition for which a bound state will exist, given that $\tau_- = t/\Delta_-$.  Setting $\Delta_-^2/t < 1$ gives
\begin{equation}
    \frac{u}{2t} \left(1-\sqrt{1-\frac{4t^2}{u^2}}\right) < 1,
\end{equation}
which always holds since we require $u > 2t$ (this can be seen because the left hand side is a decreasing function of $u/2t$ and equals $1$ when $u = 2t$); this implies the existence of an additional bound state at zero energy.  This is the discrete analogue of the Jackiw-Rebbi zero mode associated with the kink soliton solution.

Next, we may iterate this process to determine other reflectionless kink potentials with a given spectrum of bound states.  One subtlety we must observe is that we cannot guarantee that the supersymmetry process can be iterated indefinitely; if the partial continued fraction expression for some $\Delta_N^2$ is negative at a given iteration, then there does not exist a solution to Eqns. \eqref{SUSY hierarchy} and \eqref{SUSY hierarchy2}.  In the general case, we may write
\begin{equation}\label{superpotcontfrac}
   \Delta_{-N}^2 = \cfrac{t_{-N+1}^2}{u_{-N+1} -\cfrac{t_{-N+2}^2}{u_{-N+2} -\cfrac{t_{-N+3}^2}{\ddots - \cfrac{t_{0}^2}{\Delta^2}}}} \hspace{1.1cm}  \Delta_{N}^2 = u_N - \cfrac{t_N^2}{u_{N-1} -\cfrac{t_{N-1}^2}{u_{N-2} -\cfrac{t_{N-2}^2}{\ddots - \Delta^2}}}
\end{equation}
Alternatively, these continued fractions can be written in terms of the ratio of determinants of tridiagonal matrices:
\begin{equation}
    \Delta_{-N}^2 = \frac{\text{det}\begin{pmatrix}
    u_{N} & t_{N}   & \dots     & 0  & 0 \\
    t_{N}  & u_{N-1}       & t_{N-1}   & \dots     & 0   \\
     \vdots &       t_{N-1}  & \ddots         & t_{-2}    & \vdots    \\
     0 &        \vdots &          t_{-2} &     u_{-1}      & t_{-1}   \\
     0 &      0   &        \dots   &       t_{-1}    & u_0 - \Delta^2
  \end{pmatrix}}{\text{det}\begin{pmatrix}
    u_{N-1} & t_{N-1}   & \dots     & 0  & 0 \\
    t_{N-1}  & u_{N-2}       & t_{N-2}   & \dots     & 0   \\
     \vdots &       t_{N-2}  & \ddots         & t_{-2}    & \vdots    \\
     0 &        \vdots &          t_{-2} &     u_{-1}      & t_{-1}   \\
     0 &      0   &        \dots   &       t_{-1}    & u_0 - \Delta^2
  \end{pmatrix}}
\hspace{1cm}
    \frac{t_{-N+1}^2}{\Delta_{-N}^2} = \frac{\text{det}\begin{pmatrix}
    u_{-N+1} & t_{-N+2}   & \dots     & 0  & 0 \\
    t_{-N+2}  & u_{-N+2}       & t_{-N+3}   & \dots     & 0   \\
     \vdots &       t_{-N+3}  & \ddots         & t_{-1}    & \vdots    \\
     0 &        \vdots &          t_{-1} &     u_{-1}      & t_{0}   \\
     0 &      0   &        \dots   &       t_{0}    & \Delta^2
  \end{pmatrix}}{\text{det}\begin{pmatrix}
    u_{-N+2} & t_{-N+3}   & \dots     & 0  & 0 \\
    t_{-N+3}  & u_{-N+3}       & t_{-N+4}   & \dots     & 0   \\
     \vdots &       t_{-N+4}  & \ddots         & t_{-1}    & \vdots    \\
     0 &        \vdots &          t_{-1} &     u_{-1}      & t_{0}   \\
     0 &      0   &        \dots   &       t_{0}    & \Delta^2
  \end{pmatrix}}
\end{equation}
The matrices in these equations resemble submatrices of the Hamiltonian (by submatrices we mean matrices formed by deleting some set of columns and corresponding rows in the Hamiltonian).  Because the Hamiltonian is positive semi-definite and the zero-energy bound state has nonzero support at each site, all submatrices of the Hamiltonian are positive definite.  Next, we note that adding a constant $\epsilon$ to each of the onsite potentials does not change the reflectionless property of the Hamiltonian and shifts the ground state energy to $\epsilon$.  Setting $u_N \to u_N + \epsilon$, we find that a sufficient condition for both $\Delta_{N}^2$ and $t_{-N+1}^2/\Delta_{-N}^2$ to be positive is $\epsilon + \Delta^2 - u_0 > 0$ and $\epsilon - \Delta^2 > 0$.  This implies that $u_0 - \epsilon < \Delta^2 < \epsilon$.  The resulting superpartner potential will have an additional bound state at zero energy, which does not coincide with the previous bound state now at energy $\epsilon$.  With an appropriate application of a shift at every iteration, we can generate a hierarchy of reflectionless potentials.  Note that the bound on the shift $\epsilon$ that we derived is not tight, and in practice we find a much less restrictive bound.

\subsection{Construction of kink-antikink solitons}

Thus far we have described how to construct kink solutions, with asymptotically different boundary conditions at $\pm \infty$.  These topological solitons correspond to a robust zero energy mode, which is present in one of the scattering problems and absent in the other.  However, we argued that there should also be non-topological solitons -- in the high-energy context, these correspond to fermion-antifermion bound states (and in the general case, multi-fermion bound states).  In the previous section, we constructed the bound state spectrum of multi-fermion states as a function of the fermion/anti-fermion filling, but did not construct a satisfying solution for $\Delta_n$ and $\tau_n$. Here, we discuss a method for constructing such solutions using discrete supersymmetry.

As described in the previous section, we seek a solution for $\Delta_n$ which converges to the same value at $\pm \infty$.  Given an arbitrary onsite potential $u_n$ and hopping $t_n$, we may evaluate the superpotential using the continued fraction method illustrated in Eqn. \eqref{superpotcontfrac}.  Let us focus on the case where $n > 0$, for which $\Delta_n$ converges to $\Delta_+$ as $n \to \infty$.  We may first write the continued fraction as a ratio $p_n/q_n$, where, as discussed in the previous subsection, $p_n$ can be determined via the following relation:
\begin{equation}\label{transfmat}
\begin{pmatrix}
  p_N  \\
  p_{N-1}
\end{pmatrix} = \begin{pmatrix}
  \Delta^2 & -t_1^2 \\
  1 & 0 \\
\end{pmatrix} \prod_{j=1}^N \begin{pmatrix}
  u_j & -t_{j+1}^2 \\
  1 & 0 \\
\end{pmatrix} \begin{pmatrix}
  1  \\
  0
\end{pmatrix},
\end{equation}
and $q_N$ satisfies a similar relation but with $N \to N-1$.  This transfer matrix method was used to obtain the exact solution for the kink soliton where both $u_n$ and $t_n$ were constant, but works equally well in the general case.  For large $N$, the product is dominated by a product of the asymptotic form of the transfer matrix, assuming that $u_\infty = u$ and $t_\infty = t$.  The eigenvalues of the asymptotic value of the transfer matrix are $\Delta_+$ and $\Delta_- < \Delta_+$, so for generic values of the initial condition $\Delta_0 = \Delta$, $p_{N \to \infty} \sim \Delta_+^{N}$ while $q_{N \to \infty} \sim \Delta_+^{N-1}$ -- this results in the asymptotic superpotential being $\Delta_+$.  A similar analysis shows that $\Delta_{-\infty} = \Delta_-$ and we recover a kink solution.  

However, if $\Delta$ is chosen such that the eigenvector corresponding to eigenvalue $\Delta_+$ is annihilated by the matrix product in Eqn. \eqref{transfmat} (for $N$ large), then the asymptotic superpotential will be $\Delta_-$, and corresponding soliton solution will have the same boundary conditions at $\pm \infty$ -- this corresponds to a non-topological soliton.  The eigenvector of the transfer matrix corresponding to eigenvalue $\Delta_+$ is $\bm{\lambda_+}^T = (\Delta_+,1)$, so the unique value of $\Delta$ that generates these solutions solves
\begin{equation}
 \begin{matrix}\begin{pmatrix}\Delta_\star^2 & -t_1^2\end{pmatrix}\\\mbox{}\end{matrix}
  \prod_{j=1}^\infty \begin{pmatrix}
  u_j & -t_{j+1}^2 \\
  1 & 0 \\
\end{pmatrix} \begin{pmatrix}
  \Delta_+  \\
  1
\end{pmatrix} = 0,
\end{equation}
or equivalently using the notation $\bm{e_1}^T = (1,0)$ and $\bm{e_2}^T = (0,1)$,
\begin{equation}
    \Delta_\star^2 = t_1^2 \frac{\bm{e_2}^T Q \bm{\lambda_+}}{\bm{e_1}^T Q \bm{\lambda_+}}, \hspace{0.7cm} Q = \prod_{j=1}^\infty \begin{pmatrix}
  u_j & -t_{j+1}^2 \\
  1 & 0 \\
\end{pmatrix}.
\end{equation}
The corresponding non-topological solitons are unstable to perturbations in the initial condition $\Delta_\star$ and will revert to kink solitons.  Now that a value for $\Delta_0 = \Delta_\star$ is specified, we must select an appropriate choice of $u_n$.  Clearly, this choice of $u_n$ must be reflectionless, but it also must have at least one bound state (a choice of $u_n$ with no bound states such as $u_n = u$ will yield a trivial answer where $\Delta_n$ is constant everywhere).  We have constructed an example of such a potential in the previous subsection (see Eqn. \eqref{superpartnerconst}), which is a discretized version of hyperbolic secant.  The superpartner to this potential with the initial condition $\Delta^2 = \Delta_\star^2$ will not have additional bound state at zero energy as such a state is not normalizable; thus, the superpartner will have an identical spectrum.  This describes the case where supersymmetry is broken, while supersymmetry was unbroken in the case of the kink soliton.

\begin{figure*}
\begin{tabular}{cc}
  \includegraphics[scale=0.575]{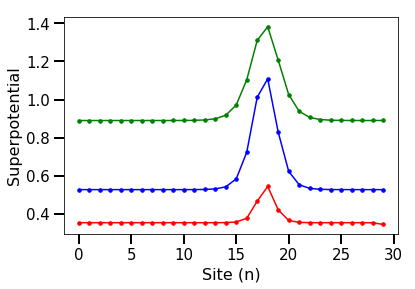} &   \includegraphics[scale=0.575]{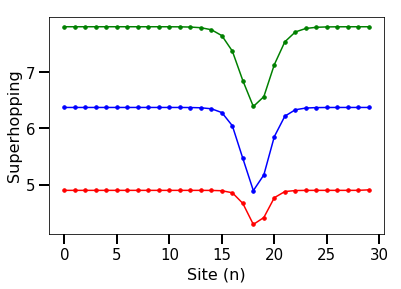} \\
\end{tabular}
\caption{A plot of the superpotential and the superhopping obtained from the discrete hyperbolic secant potential given in Eqn. \eqref{superpartnerconst} for a variety of asymptotic values of $t$ and $u$.  The boundary condition $\Delta_0^2 = \Delta_\star^2$ was chosen, which results in non-topological soliton solutions shown.  Note that the chain is cut off after 30 lattice sites due to the exponential numerical instability caused from small deviation from $\Delta_\star$.}
\label{fig:nontopsol}
\end{figure*}

We verify that the choice of potential $u_n$ and initial condition $\Delta_\star$ indeed generates non-topological solitons, as shown in Figure \ref{fig:nontopsol}.  As expected, these solitons can be viewed as a sum of a propagating kink and antikink topological soliton.  Similar kinds of non-topological solitons can be constructed for potentials with more than a single bound state by an analogous procedure.

\section{Chiral Gross-Neveu model and superconductivity}\label{chiralGNsection}

Thus far, we worked exclusively with the Gross-Neveu model, showing the correspondence between semiclassical saddle point solutions of the action and reflectionless potentials of the effective scattering problem.  We were able to explicitly retrieve these reflectionless solutions using discrete supersymmetry.  We analyzed both topological and non-topological soliton solutions, the former of which appears as a kink in the magnitude of the real scalar field $\Delta$, and the latter of which occurs due to the merging of kink and antikink solutions.  In superconductors, the order parameter is complex, and kink solitons can become dynamical.  To reproduce this case, we consider a modification of the Gross-Neveu model, the chiral Gross-Neveu model (equivalently the Nambu-Jona-Lasinio model), which has an additional chiral term necessary for the order paraneter to be complex.  In this section, we perform a similar analysis of this model, and construct nontrivial soliton solutions in lattice superconductors, relevant for cold atom experimental platforms.

The chiral Gross-Neveu model, as first discussed by Shei \cite{Shei_PRD_1976}, is a modification of the Gross-Neveu model that preserves chiral symmetry $\psi \to e^{i \theta \gamma_5} \psi$:
\begin{equation}
    \mathcal{L} = \sum_{k=1}^{N}i\overline{\psi}^{(k)}\slashed{\partial}\psi^{(k)} +\frac{g^2}{2}\left(\sum_{k=1}^{N}\overline{\psi}^{(k)}\psi^{(k)}\right)^2 - \frac{g^2}{2}\left(\sum_{k=1}^{N}\overline{\psi}^{(k)} \gamma_5 \psi^{(k)}\right)^2.
\end{equation}
As in the rest of the text, we use a representation of the gamma matrices where $\gamma_0 = -\sigma_x$, $\gamma_1 = -i \sigma_z$, and $\gamma_5 = \gamma_0 \gamma_1 = \sigma_y$.  As defined on a lattice, the chiral Gross-Neveu model is
\begin{equation}
    \mathcal{L}_n =  \sum_{k=1}^{N}\Psi_{n}^{\dagger(k)}(t)\begin{pmatrix} i\frac{\partial}{\partial t} & \partial_{n} \\ \partial_{n}^{\dagger} & i\frac{\partial}{\partial t} \end{pmatrix} \Psi_{n}^{(k)}(t)
    +\frac{g^2}{2}\left(\Psi_{n}^{\dagger (k)}(t)\gamma_0\Psi_{n}^{(k)}(t)\right)^2 - \frac{g^2}{2}\left(\Psi_{n}^{\dagger (k)}(t)\gamma_0 \gamma_5\Psi_{n}^{(k)}(t)\right)^2.
\end{equation}
We will follow Shei's analysis and determine the saddle point solutions of this lattice action.  In the continuum limit, the chiral GN model has saddle point equations which exactly corresponds to the BdG equation and self-consistency condition in superconductors and fermionic superfluids; the Lagrangian above therefore corresponds to analogous systems defined on a lattice.

\subsection{Action in terms of scattering data}
To map the action onto a scattering problem, we require the introduction of two Hubbard-Stratonovich fields, which we will call $\chi$ and $\eta$ with the full order parameter $\Delta = \chi + i \eta$; suppressing the sum over flavors, the Lagrangian then becomes
\begin{equation}
    \mathcal{L}_{n} = \Psi_{n}^{\dagger} \begin{pmatrix} i\frac{\partial}{\partial t} - g\eta_n(t) & \partial_{n}+g\chi_n(t) \\ \partial_{n}^{\dagger}+g\chi_{n}(t) & i\frac{\partial}{\partial t} + g\eta_n(t) \end{pmatrix} \Psi_{n}-\frac{\chi_{n}^2(t)}{2} -\frac{\eta_{n}^2(t)}{2}. 
\end{equation}
As before, we assume that the Hubbard-Stratonovich fields are static.  Then, the action can be split into two pieces as in the case of the Gross-Neveu model:
\begin{align}
    S_1(\Delta) &= -\frac{T}{2} \sum_{n} (\chi_n^2 + \eta_n^2),\\
    S_2(\Psi, \Delta) &=  \sum_{n}\int_{0}^{T} dt\, \Psi_{n}^{\dagger} \begin{pmatrix} i\frac{\partial}{\partial t} - g\eta_n & \partial_{n}+g\chi_{n} \\ \partial_{n}^{\dagger}+g\chi_{n} & i\frac{\partial}{\partial t} + g\eta_n \end{pmatrix} \Psi_{n}.
\end{align}
The equations of motion now correspond to the following eigenvalue problem (assuming simple oscillatory time dependence for the fermionic fields)
\begin{equation}
    \begin{pmatrix}- g\eta_n & \partial_{n}+g\chi_n \\ \partial_{n}^{\dagger}+g\chi_n & g\eta_n \end{pmatrix} \Psi_{n} = \omega \Psi_{n}.
\end{equation}
Unlike in the original Gross-Neveu model, this equation cannot be factorized into two independent Schr\"odinger equations for each component of the fermionic field due to the presence of the chiral term.  As a result, we must re-derive trace identities for the above Dirac operator and relate the action to suitably defined scattering data.  As discussed in Appendix \ref{Appendix A}, we may define the Jost functions for this scattering problem, which are written as $f_{\pm}$ and $g_{\pm}$ -- the subscript $\pm$ is present due to the two-sheet structure of the Dirac equation, which supports both positive and negative energy modes.  In the Dirac equation, the positive and negative sheets are decoupled, and so the action can be written in terms of two reflection coefficients $r_+(\phi)$ and $r_-(\phi)$, each of which corresponds to a scattering process involving the positive or negative modes separately.  As in the Gross-Neveu model, we assume that both $a_+(\phi)$ and $a_-(\phi)$ have the same structure as in Eqns. \eqref{scattering a} and \eqref{bound a}.  Therefore, by the Poisson-Jensen formula, we may write
\begin{align}
    \log{|a_+(z)|} + &\log{|a_-(z)|} = \sum_{j=1}^{M_+}\log {\left|\frac{z-z_{j,+}}{1-z_{j,+}z}\right|} + \sum_{j=1}^{M_-}\log {\left|\frac{z-z_{j,-}}{1-z_{j,-}z}\right|} \nonumber \\&-\frac{1}{4\pi} \int_{0}^{2\pi}\Re\left( \frac{e^{i\phi}+z}{e^{i\phi}-z}\right)\left[\log\left(1-|r_+(e^{i\phi})|^2\right) + \log\left(1-|r_-(e^{i\phi})|^2\right )\right]\,d\phi,
\end{align}
where $M_{\pm}$ are the number of bound states in the scattering problems defined on the positive and negative sheets.  On a technical note, we show in Appendix A that $a_+(z)$ and $a_-(z)$ are separately not meromorphic in the unit disk, but their product is; thus, the Poisson-Jensen formula must be applied to $\log (a_+(z) a_-(z))$.  As a result, the left hand side of the above equation admits a Laurent series expansion as derived in Appendix \ref{Appendix A}.  In particular, like for the Schrodinger equation, we show that
\begin{equation}
    -\log \frac{a_+(z)}{a_+(0)} - \log \frac{a_-(z)}{a_-(0)} = h_+(z) + h_-(z) + \sum_{k = 0}^\infty \frac{\text{Tr}(H^{2k})}{\omega^{2k} k},
\end{equation}
where $h_{\pm}$ are unimportant functions and 
\begin{equation}
    \omega^2(z) = \eta_{\infty}^2 + \left( \frac{\tau_{\infty}(z + z^{-1})}{2} +\chi_{\infty}\right)^2 -  \frac{\tau_{\infty}^2(z - z^{-1})^2}{4}.
\end{equation}
and the matrix $H$ is the Dirac operator:
\begin{equation}\label{diracop}
    H = \begin{pmatrix}- g\eta_n & \partial_{n}+g\chi_n \\ \partial_{n}^{\dagger}+g\chi_n & g\eta_n \end{pmatrix}.
\end{equation}
Henceforth, we shall set $g = 1$.  As in the case of the Gross-Neveu model, we match the series expansion obtained via the trace identity with that obtained via the Poisson-Jensen formula.  Matching to lowest order in $z$ gives the first nontrivial trace identity:
\begin{align}
    -\frac{1}{2\pi} \int_{0}^{2\pi}&\left[\log\left(1-|r_+(e^{i\phi})|^2\right) +\log\left(1-|r_-(e^{i\phi})|^2\right) \right]\cos{(\phi)}\,d\phi  +\sum_{j=1}^{M_+}\left(z_{j,+}-\frac{1}{z_{j,+}}\right)+ \nonumber \\&\sum_{j=1}^{M_-}\left(z_{j,-}-\frac{1}{z_{j,-}}\right) = C(N, \chi_\infty, \eta_\infty, \tau_\infty) + \frac{4}{\tau_\infty\chi_\infty}\sum_n (\eta_n^2 + \chi_n^2 + \tau_n^2),
\end{align}
where $C$ is a constant that can be eliminated by an appropriate shift of the vacuum energy and will be ignored henceforth.  Here, we have used the fact that $\text{Tr} (H^2) = \sum_n (\eta_n^2 + \chi_n^2 + \tau_n^2)$.  Using the assumption that $\tau_\infty \chi_\infty = 1$ and assuming that the sum of the squared values of $\tau_i$ are bounded and fixed, the second term in the action can be written in terms of the scattering data:
\begin{align}
    S_1 = -\frac{T}{2} \sum_n (\chi_{n}^2 + \eta_{n}^2) = \frac{T}{2\pi} \int_{0}^{2\pi}&\left[\log\left(1-|r_+(e^{i\phi})|^2\right) +\log\left(1-|r_-(e^{i\phi})|^2\right) \right]\cos{(\phi)}\,d\phi \nonumber\\ &+\sum_{j=1}^{M_+}\left(z_{j,+}-\frac{1}{z_{j,+}}\right)+ \sum_{j=1}^{M_-}\left(z_{j,-}-\frac{1}{z_{j,-}}\right).
\end{align}
Next, we must write the integration over the fermionic variables in terms of the scattering data.  This can be done in a very similar way to what was done in the Gross-Neveu model.  First, Eqn. \eqref{value of path integral} still holds.  However, in this case we must distinguish the positive and negative sheets because the corresponding bound state spectra may not coincide.  In the Gross-Neveu case, this was not necessary because the theory was invariant under charge conjugation.  Therefore, we write
\begin{align}
     \mathcal{I}_{1}(\chi, \eta) &= \mathcal{I}_{1}(0) \prod_{k}(e^{i\alpha_{+,k}/2}+e^{-i\alpha_{+,k}/2}) \prod_{k'}(e^{i\alpha_{-,k'}/2}+e^{-i\alpha_{-,k'}/2}) \\
     &= \mathcal{I}_{1}(0)\exp\left(\frac{i}{2}\sum_{k}|\alpha_{+,k}| + |\alpha_{-,k}| \right)\prod_{\alpha_{+,k} > 0}(1+e^{-i\alpha_{+,k}}) \prod_{\alpha_{-,k} > 0}(1+e^{-i\alpha_{-,k}}) \\
     &= \mathcal{I}_{1}(0)\exp\left(\frac{i}{2}\sum_{k}|\alpha_{+,k}| + |\alpha_{-,k}| \right)\sum_{\{n_i, m_j\}} e^{-i\sum_{\alpha_{+,i}>0} n_i\alpha_{+,i} -i\sum_{\alpha_{-,j}>0} m_j\alpha_{-,j}} \prod_{i, j} q_{n_i} q_{m_j},
\end{align}
where the combinatorial coefficients are defined without an additional factor of two, $q_{n_i} = \frac{N!}{n_i!(N-n_i)!}$.  Then, the full effective action can be written as
\begin{equation}\label{chiralGNscat}
    S_{\text{eff}} = S_1 + \frac{N}{2} \sum_{k}(|\alpha_{+,k}| + |\alpha_{-,k}|) - \sum_{k} n_{+,k}|\alpha_{+,k}| - \sum_{k} n_{-,k}|\alpha_{-,k}| + \text{const.}
\end{equation}
Next, we make the assumption that there is only a single discrete bound state for each branch: that is, $n_{+,0}$ fermions occupy mode $\alpha_{+,0}$ and $n_{-,0}$ fermions occupy mode $\alpha_{-,0}$ (the rest fill the continuum Fermi sea).  Next, we need to relate the second term in the equation above as a function of the scattering data.  Again, note that in the time independent case, the Floquet indices are $T$ times the corresponding energies.  Because the positive and negative branches are independent from one another, the associated scattering problems are decoupled, and a similar analysis to that presented in Appendix \ref{Appendix B} gives (up to an infinite constant that is corrected by a suitable renormalization of the vacuum energy)
\begin{equation}
    \sum_{k}(|\alpha_{+,k}| + |\alpha_{-,k}|) = 2T \int_0^{2\pi} \frac{d\phi}{2 \pi} \frac{\sin \phi}{\sqrt{\eta_0^2 + \chi_0^2 + \tau_0^2 - \cos \phi}} (\delta_+(\phi) + \delta_-(\phi)),
\end{equation}
where $\eta_0^2 + \chi_0^2 + \tau_0^2$ is the value of $\eta_n^2 + \chi_n^2 + \tau_n^2$ at infinity.  Using the imaginary part of the Jensen formula presented in Eqn. \eqref{JensenIm}, the action in Eqn. \eqref{chiralGNscat} can be written in terms of the reflection coefficient as well as the energy of the bound state
\begin{align}\label{chiralGNscatfull}
    \frac{S_{\text{eff}}}{T} = \frac{1}{2\pi} \int_{0}^{2\pi}&\left[\log\left(1-|r_+(e^{i\phi})|^2\right) +\log\left(1-|r_-(e^{i\phi})|^2\right) \right]\cos{(\phi)}\,d\phi \nonumber\\ &+\sum_{j=1}^{M_+}\left(z_{j,+}-\frac{1}{z_{j,+}}\right)+ \sum_{j=1}^{M_-}\left(z_{j,-}-\frac{1}{z_{j,-}}\right) - n_{+,0}|\omega_{+,0}| - n_{-,0}|\omega_{-,0}| \nonumber\\ &+ N \int_0^{2\pi} \frac{d\phi}{2 \pi} \frac{\sin \phi}{\sqrt{\eta_0^2 + \chi_0^2 + \tau_0^2 - \cos \phi}} (\delta_+(\phi) + \delta_-(\phi)).
\end{align}

The saddle point corresponding to this action implies that the effective scattering problems for both the positive and negative sheets are reflectionless.  Furthermore, the corresponding bound state energy can be constructed by finding a minimum of the action with respect to $z_{j, \pm}$.  The corresponding bound state energies are similar to those computed in the case of the Gross-Neveu model, and like before, we have two types of soliton solutions, topological and nontopological solitons.  However, because the HS parameter is complex, there is a richer structure of saddle point solutions with nontrivial phase slip.
\subsection{Soliton solutions}

We now proceed to find saddle point solutions corresponding to the action in Eqn. \eqref{chiralGNscatfull}.  As the saddle point corresponds to reflectionless potentials in both of the scattering processes in the Dirac equation, we need to utilize the inverse scattering approach to retrieve the most general form of such a solution, which we will not attempt to pursue here.  One such class of solutions corresponds to $\eta_n = 0$ and $\chi_n$ a reflectionless superpotential; in this case, the chiral Gross-Neveu model reduces to the Gross-Neveu model and our previous analysis applies.  These solutions are likely to be of most relevance for experimental verification.

There is another saddle point solution which can be achieved through minimal additional effort.  We square the Dirac operator $H$ defined in Eqn. \eqref{diracop} to obtain
\begin{equation}
    H^2 = \begin{pmatrix} \eta_n^2 + (\partial_{n}+\chi_n)(\partial_{n}^{\dagger}+\chi_{n}) & \overline{\partial}_n \\ \overline{\partial}_n^{\dagger} & \eta_n^2 + (\partial_{n}^{\dagger}+\chi_{n})(\partial_{n}+\chi_n) \end{pmatrix},
\end{equation}
where $\overline{\partial}_n f_n = \tau_{n+1}(\eta_{n+1} - \eta_{n}) f_{n+1}$.  If $\eta_n = \eta$ is a constant, then the off-diagonal blocks vanish and the operator has a supersymmetric structure apart from a shift $\eta^2$, which does not affect the energy eigenstates.  The eigenstates of the square of the Dirac operator are those of the original Dirac operator, and if $\chi$ is chosen to be a reflectionless superpotential, then the eigenstates of $H^2$ are reflectionless.  This implies that both the positive and negative scattering problems are reflectionless for the Dirac operator, therefore satisfying the saddle point condition.

For these solutions, the Hubbard-Stratonovich field has a non-zero albeit constant imaginary part.  If $\chi$ is a non-topological soliton, then it may be expressed as a kink-antikink pair as was previously deduced (assuming occupation of only a single bound state).  If $\chi$ is a topological kink soliton, then the Hubbard-Stratonovich field undergoes a phase slip by a nontrivial angle $\theta$, given by
\begin{equation}
    \tan \theta = -\frac{\eta (\chi_+ - \chi_-)}{\chi_+\chi_- + \eta^2} = - \frac{\eta \sqrt{u}}{t + \eta^2} \left[\left(\frac{1}{2} + \frac{1}{2}\sqrt{1 - \frac{4 t^2}{u^2}}\right)^{1/2} - \left(\frac{1}{2} - \frac{1}{2}\sqrt{1 - \frac{4 t^2}{u^2}}\right)^{1/2}\right],
\end{equation}
where $\chi_\pm$ are given in Eqn. \eqref{suppm}.  In the continuum case, it has been shown that such solitons experience dynamics if the phase slip does not equal $\pi$, and one would need to solve for time-dependent saddle point solutions.  The velocity of propagation of the soliton strongly depends on the phase angle: as the propagation velocity increases, the phase angle decreases towards zero and vanishes at a critical velocity, at which the soliton becomes unstable in accordance to the Landau criterion \cite{Efimkin_PRA_2015}.  The theory of a dynamical or moving soliton has not been pursued here, though the authors speculate that the time dependence simply can be constructed from replacing $n$ with $n - v t$ in expressions for $\chi_n$, $\eta_n$, and $\tau_n$.

Finally, constructing reflectionless potentials of the Dirac equation is a rich problem, and there is no elegant method one may use to construct such potentials via supersymmetry, unlike in the case of the Schr\"odinger equation.  Classification of such soliton solutions would be a reasonable next step to pursue \cite{Nogami, Thies_2013}.

\section{Relation to the Toda hierarchy}

In Section \ref{Saddle point} we discussed the KdV equation and the KdV hierarchy, which is a series of nonlinear differential equations that are integrable and possess soliton solutions. To construct the hierarchy, one considers each trace identity of the Schr\"odinger equation as a Hamiltonian and constructs the corresponding equations of motion in terms of the scattering data.  As has been emphasized, the soliton solutions of the KdV equation are related to those appearing as saddle point solutions in the Gross-Neveu model.

In the discrete case, the lattice Gross-Neveu model turns out to analogously possess resemblance to the Toda hierarchy, a series of integrable dynamical systems.  To illustrate the resemblance, we reconstruct the Toda lattice Hamiltonian by utilizing the discrete trace identities; the method that we discuss had been first developed by Faddeev \cite{Faddeev_2007_Springer}.  However, there are two differences from the continuous case that we must reconcile.  First, for discrete systems each trace identity given by $K_n$ (via Eqn. \eqref{Trace Identities}) cannot be treated as a Hamiltonian system because the dynamics induced by such a system does not preserve the phase space measure. Instead, Faddeev has shown that one needs to consider the combination $K_n-K_{n-2}$ as the Hamiltonian. Second, unlike in the case of the KdV equation, both the potential and hopping are considered dynamical variables, and are related to the original canonically conjugate position and momentum variables of the Toda lattice by a Flaschka transformation. The first Hamiltonian of the hierarchy is
\begin{equation} \label{Toda hierarchy 1}
    H_1 = K_2 = \sum_n u_{n}^2 + 2 t_n^2,
\end{equation}
where $u_n$ is the potential and $t_n$ is the hopping. If we identify them as the Flaschka variables defined as $u_n = -\frac{p_n}{2}$ and $t_n = \frac{1}{2}e^{-\frac{(q_{n+1}-q_{n})}{2}}$ then the Hamiltonian corresponds to the Toda lattice \cite{Toda_Springer_1989}. The equations of motion of such a system therefore preserves the magnitude of the reflection coefficient, as observed in the KdV equation; soliton solutions correspond to time-dependent reflectionless potentials.  Continuing in a similar manner generates all Hamiltonians in the Toda hierarchy.  

One subtlety that we have ignored is the fact that the equations of motion of the Gross-Neveu model are those of a massless Dirac particle, which decouples into two Schr\"odinger equations.  Thus, it is more accurate to draw the analogy between the Gross-Neveu model and two interrelated Toda lattices.  The soliton solutions to an analogous dynamical system will correspond to propagating kinks or domain walls.  To obtain this dynamical system, we utilize the Lax pair formalism: we construct two operators $L$ (which is antisymmetric) and $H$ such that $\dot{H} = [L,H]$.  It is simple to show that this dynamical system has an infinite set of conserved quantities $K_p = \text{Tr}(H^p)$, implying its integrability.  We choose the operator $H$ to be the Dirac Hamiltonian in Eqn. \eqref{diracop} with $\eta_n = 0$ and $\chi_n$ relabelled to be $\Delta_n$.  Therefore, the conserved quantities coincide with the right hand sides of the trace identities. We also choose the following ansatz for the block diagonal matrix $L = \text{diag}(h,h')$, where $h$ and $h'$ are the matrices with nonzero matrix elements $h'_{n,n+1} = -h'_{n+1,n} = \Delta_n \tau_{n+1}$ and $h_{n,n+1} = -h_{n+1,n} = \Delta_{n+1} \tau_{n+1}$.  With this choice, $H$ and $L$ form a Lax pair, resulting in the dynamical system
\begin{equation}
    \frac{d \Delta_n}{d t} = \Delta_n\left(\tau_n^2 - \tau_{n+1}^2\right) \hspace{0.8cm} \frac{d \tau_n}{d t} = \tau_n\left(\Delta_{n-1}^2 - \Delta_{n}^2\right).
\end{equation}
If we perform a Flaschka transformation $\Delta_n = e^{\frac{1}{2}(q_{n+1} - q_n)}$ and $\tau_n = e^{p_n}$, the equations of motion describe a Hamiltonian system with the Hamiltonian
\begin{equation}
    H_1 = K_1 = \sum_n \tau_n^2 + \Delta_n^2 = \sum_n e^{2 p_n} + \sum_n e^{q_{n+1} - q_n},
\end{equation}
which is similar to the Toda lattice Hamiltonian but with the quadratic kinetic energy replaced by the exponential of the momentum operator. For the case of the chiral Gross-Neveu model (with $\eta_n \neq 0$), the authors have no found a satisfying Lax pair $L$ so we cannot deduce dynamical systems with complex soliton solutions.  However, for a Dirac operator with a mass term,
\begin{equation}
    H = \begin{pmatrix}- g\eta_n & \partial_{n}+g\chi_n \\ \partial_{n}^{\dagger}+g\chi_n & g\zeta_n \end{pmatrix},
\end{equation}
there exists a Lax pair matrix $L$ with a similar block matrix structure, but with non-trivial off-diagonal blocks having the same matrix structure of $\partial_n$.  We will not write down the associated dynamical system, as it possesses a different class of soliton solutions that we have not constructed.

\begin{comment}
The next Hamiltonian in the hierarchy is
\begin{equation} \label{Toda hierarchy 2}
    H_2 = K_3-K_1 = \sum_n \left[\frac{u_n^3}{3}+u_n\left(t_{n-1}^2+t_n^2\right)-2u_n\right]
\end{equation}
Applying the Flaschka transformation once more and identifying $p_n$ and $q_n$ as canonically conjugate variables, the corresponding equations of motion are
\begin{align}
    \dot{u}_n &= \frac{t_n^2}{2}\left(u_n+u_{n+1}\right) - \frac{t_{n-1}^2}{2}\left(u_n+u_{n-1}\right) \\
    \dot{t}_n &= \frac{t_n}{4}(u_{n+1}^2-u_{n}^2)+\frac{t_n}{4}(t_{n+1}^2-t_{n-1}^2).
\end{align}
\end{comment}

\begin{comment}
One can construct the other dynamical systems in the hierarchy by utilizing $\dot{H}_p = [L,H_p]$ with $H_p = H^p$ with the same Lax pair $H$ and $L$.  Using the same Flaschka transformation, we may write, for example, the next Hamiltonian in the hierarchy as
\begin{align}
    H_2 &= K_2 = \sum_n \tau_n^4 + \Delta_n^4 + 2 \Delta_n^2(\tau_n^2 + \tau_{n+1}^2) \nonumber \\ &= \sum_n e^{4 p_n} +  e^{2(q_{n+1} - q_n)} + 2 e^{2 p_n + q_{n+1} - q_n} + 2 e^{2 p_n + q_{n+2} - q_{n+1}}.
\end{align}
\end{comment}

\section{Conclusions}

In this paper, we describe a general formalism to construct soliton solutions corresponding to saddle points of a lattice quantum field theory, which is based on the semiclassical method developed by Dashen, Hasslacher, and Neveu \cite{Neveu_PRD_1975, Shei_PRD_1976}.  We first focus our attention to a lattice generalization of the Gross-Neveu model with $N$ flavors of fermions.  The equations of motion decouple into two discrete Schr\"odinger equations, and utilizing trace identities for the Schr\"odinger equation, we rewrite the action in terms of scattering variables -- namely, the reflection coefficient and the bound state spectrum.  The saddle point solutions to this action correspond to the Hubbard-Stratonovich parameter being a reflectionless potential.  Depending on the boundary conditions at $\pm \infty$, we derive the bound state spectrum for either topological or non-topoogical solitons, the former of which possesses a robust zero mode.  To construct these potentials explicitly, we develop a generalization of discrete supersymmetry, in which we create a hierarchy of reflectionless tight-binding Hamiltonians with a bound state spectrum of our choice.  We are able to retrieve both kink solitons and kink-antikink solitons using this formalism.  Both such solitons are described by a site-dependent modulation in the onsite potential and the hopping, the latter of which is absent in the continuum limit.

We then proceed to applying the semiclassical method to the lattice chiral Gross-Neveu model.  In doing so, we derive a set of trace identities for the Dirac operator, and show that the action can be written in terms of scattering data of the scattering problems of the positive and negative energy modes.  In a similar way, the particle spectrum can be obtained, and saddle point solutions correspond to reflectionless potentials of the Dirac equation.  We write down one such example, a kink soliton undergoing a non-$\pi$ phase slip; these solitons likely have nontrivial dynamics that we have not accounted for.  Finally, we relate the soliton solutions obtained from the saddle points to exact solitons in the Toda lattice.  In the process, we write down dynamical systems whose solutions are propagating domain walls/kinks, in analogy to the static solutions derived in the field theories.

There are multiple generalizations that are straightforward to consider given our formalism.  First, we have exclusively considered a single-band model.  To construct a two-band (and in general, a multi-band) model, we simply set asymptotic conditions where $\Delta_{2n+1} = \Delta_1^{(\pm)}$ and $\Delta_{2n} = \Delta_2^{(\pm)}$ as $n \to \pm\infty$, which is in the spirit of the Su-Schrieffer-Heeger model \cite{SSH, Campbell_PRB_1981}.  The action can be written in terms of scattering data by an appropriate modification of trace identities for the Schr\"odinger operator with staggered boundary conditions.  Reflectionless soliton solutions can be computed easily using the discrete supersymmetry formalism.  Another possible extension would be to apply a magnetic field, so that the hopping undergoes a phase slip from $-\infty$ to $\infty$.  Second, it may be possible to pursue a similar analysis with long-range hopping.  In this case, there may be multiple scatterers at a given energy, and the scattering data will consist of a potentially large scattering matrix.  Depending on the structure of the hopping, discrete supersymmetry may still provide a means for obtaining classes of solutions with scattering properties prescribed by the saddle point solution.  Finally, it is interesting to pursue an analogous analysis in 2+1 dimensions, where the inverse scattering method can still be formulated, though supersymmetry may not be a useful tool for extracting soliton solutions.

The original motivation for studying both the Gross-Neveu and the chiral Gross-Neveu model is that both models (especially the latter) are equivalent to the self-consistent Bogoliubov de-Genne Hamiltonians describing superconductors and/or fermionic superfluids \cite{Takahashi_PRL_2013}.  As a result, it is possible that lattice soliton solutions presented in the text can be observed in cold atom experiments \cite{Yefsah_Nature_2013}.  A more complete analysis should also include an analysis of the dynamics of such soliton solutions, where it is known that the velocity of propagation is related to the phase slip of a kink soliton \cite{Efimkin_PRA_2015}.  Furthermore, the kink solitons that were presented for the chiral Gross-Neveu model are not the only family of reflectionless potentials; a complete classification of transparent potentials of the Dirac equation would be of importance to understanding the full saddle point structure of the action \cite{Nogami, Thies_2013}.

Finally, discrete supersymmetry is itself a rich topic with many unexplored questions.  Though the application of the methodology in this paper was for obtaining soliton solutions, discrete symmetry can be applied to other kinds of 1D and quasi 1D systems to explore potentially interesting physics.  A complete classification of what kinds of systems are closed under a discrete SUSY transformation has not been performed.  Discrete SUSY may also have significance when applied to non-Hermitian or time-reversal symmetry breaking Hamiltonians.  A potential application of discrete SUSY would be to probe topological properties of a system.  Since a single iteration of supersymmetry removes or adds a zero mode, the topological (Witten) index of a system can itself be altered -- it would be interesting to pursue an analysis in this direction \cite{queralto2020topological}.

\appendix
\section{Discrete trace identities}
\label{Appendix A}
\subsection{Schr\"odinger equation}
The following is an abridged version of the analysis presented in Toda's book \cite{Toda_Springer_1989}.  In this appendix we will derive Eqn. \eqref{connceting a with potential}.  The key arguments in this derivation rely on relating the trace of the Green's function to both the scattering data as well as to the explicit onsite potentials and hoppings.  We know that the Jost functions $f_n(z)$ and $g_n(z)$ introduced in the main text of the paper satisfy the discrete Schr{\"o}dinger equation; assuming the onsite potential has an asymptotic value $u$ and the hoppings have an asymptotic value $1$,
\begin{align}\label{eqn for f}
    -t_{n+1}f_{n+1}\left(z\right)-t_{n}f_{n-1}\left(z\right) + u_nf_n\left(z\right)  &=\left(u-z-\frac{1}{z}\right)f_n(z),  \\
    \label{eqn for g}
    -t_{n+1}g_{n+1}(z)-t_{n}g_{n-1}(z)+ u_ng_n(z) &= \left(u-z-\frac{1}{z}\right)g_n(z).  
\end{align}
where we have used the energy dispersion $\omega = u-(z+1/z)$. We multiply Eqn. \eqref{eqn for f} and  \eqref{eqn for g} by $g_{n}(z)$ and $f_n(z)$ respectively. Subtracting these equations, we find
\begin{equation} \label{eqn for f and g}
    t_{n+1}\left(g_{n+1}(z)f_n(z)-g_{n}(z)f_{n+1}(z)\right) = t_{n}\left(g_{n}(z)f_{n-1}(z)-g_{n-1}(z)f_{n}(z)\right).
\end{equation}
From this, we can identify a quantity (a discrete version of the Wronskian of $f_n(z)$ and $g_n(z)$) that is conserved for all $n$:
\begin{equation} \label{wronskian}
    w = t_{n+1}\left(g_{n+1}(z)f_n(z)-g_{n}(z)f_{n+1}(z)\right).
\end{equation}
We note that the conserved quantity $w$ is merely a disrete generalization of the Wronskian of $f_n(z)$ and $g_n(z)$. We can find the absolute value of $w$ in the asymptotic limit:
\begin{equation}
    w =\lim_{n \to -\infty} t_{n+1}\left(g_{n+1}(z)f_n(z)-g_{n}(z)f_{n+1}(z)\right) = a(z)\left(\frac{1}{z}-z\right).
\end{equation}
%We can use Eqn. \eqref{eqn for g} to write the Wronskian as:
%\begin{multline} \label{Wronskian form 2}
%    w = \left(-t_{n}g_{n-1}(z)+(2t_{n} + u_n)g_n(z)\right)f_{n}(z) \\  %+\left(z+\frac{1}{z}-2\right)g_n(z) f_n(z)-t_{n+1}g_{n}(z)f_{n+1}(z).
%\end{multline}
Now we define a matrix $G_{n,m}$ with elements
\begin{align*}
    G_{n,m}(z) &= \frac{f_n(z)g_m(z)}{w} \text{ when } n\geq m \\
    G_{m,n}(z) &= G_{n,m}(z).
\end{align*}
%In terms of $G$ we can write Eqn. \eqref{Wronskian form 2} as:
%\begin{equation} \label{first eqn for G}
%    -t_{n+1}G_{n+1,n} + (u_n+2t_{n}-2)G_{n,n} - t_{n}G_{n-1,n} + \left(z+\frac{1}{z}\right)G_{n,n} = 1.
%\end{equation}
%Multiplying Eqn. \eqref{eqn for f} with $g_m(z)$ and Eqn. \eqref{eqn for g} with $f_m(z)$ we get that when $n \neq m$ the function $G$ satisfies:
%\begin{equation} \label{second eqn for G}
%    -t_{n+1}G_{n+1,m} + (u_n+2t_n-2)G_{n,m} - t_{n}G_{n-1,m} + %\left(z+\frac{1}{z}\right)G_{n,m} = 0 .
%\end{equation}
%Combining Eqn. \eqref{first eqn for G} and Eqn. \eqref{second eqn for G} we get:
%\begin{equation} \label{final eqn for G}
%    -t_{n+1}G_{n+1,m} + (u_n+2t_n-2)G_{n,m} - t_{n}G_{n-1,m} + \left(z+\frac{1}{z}\right)G_{n,m} = 0.
%\end{equation}
One can show by explicit computation that $G$ satisfies
\begin{equation} \label{G in matrix form}
    \left[H+\left(z+\frac{1}{z} - u\right)\mathbb{1}\right]G = \mathbb{1},
\end{equation}
where $H$ is the Hamiltonian with matrix elements $H_{n,n-1}=-t_{n}, H_{n,n+1}=-t_{n+1}$ and  $H_{n,n} = u_n$. Note that this equation identifies $G$ as the Green's function for the operator $H+(z+\frac{1}{z})$. Using Eqn. \eqref{G in matrix form} one can write $G$ in terms of the potential. We return to Eqn. \eqref{eqn for f} and take a derivative with respect to $z$ to get
\begin{equation} \label{derivative of f}
    -t_{n+1}\dot{f}_{n+1}(z)-t_{n}\dot{f}_{n-1}(z) + u_n\dot{f}_n(z) + \left(z+\frac{1}{z} - u\right)\dot{f}_n(z) + \left(1-\frac{1}{z^2}\right)f_n(z)=0.
\end{equation}
Here we have used the notation $\dot{f}(z) \equiv \frac{df(z)}{dz}$. After some algebraic manipulations of the above equation, one can show
%We multiply Eqn. \eqref{derivative of f} by $g_n(z)$ and Eqn. \eqref{eqn for g} by $\dot{f}_n(z)$. Then we subtract the two equations to get:
%\begin{equation} \label{fngn in terms of derivative}
%    t_{n+1}\left(g_{n+1}\dot{f}_{n}-g_{n}\dot{f}_{n+1}\right)-t_{n}\left(g_{n}\dot{f}_{n-1}-g_{n-1}\dot{f}_{n}\right) = -\left(1-\frac{1}{z^2}\right)f_ng_n.
%\end{equation}
%Define $U_n \equiv t_{n+1}\left(g_{n+1}\dot{f}_{n}-g_{n}\dot{f}_{n+1}\right)$.  This allows us to write Eqn. \eqref{fngn in terms of derivative} as a difference relation in terms of $U$:
\begin{equation} \label{eqn for U}
    -\left(1-\frac{1}{z^2}\right)f_ng_n = U_n-U_{n-1},
\end{equation}
where we have defined $U_n \equiv t_{n+1}\left(g_{n+1}\dot{f}_{n}-g_{n}\dot{f}_{n+1}\right)$. One can show that the asymptotic form of $U_n$ is given by 
\begin{equation} \label{asymptotic Un}
    U_{n} = 
  \begin{cases}
   \dot{a}(z)\left(\frac{1}{z}-z\right)+a(z)\left[\frac{n+1}{z^2}-n\right], & \text{for } n \to \infty 
    \\[10pt]
   a(z)\left[n - \frac{n-1}{z^2}\right], & \text{for } n \to -\infty
  \end{cases}.
\end{equation}
Eqn. \eqref{asymptotic Un} is valid for $|z|<1$. Using Eqn. \eqref{eqn for U} and the asymptotic form of $U_n$ we get
%\begin{equation}
%    -\left(1-\frac{1}{z^2}\right)\sum_{n = -N+1}^Nf_ng_n = U_N-U_{-N}.
%\end{equation}
%For $N\gg 1$ using the asymptotic form of $U_n$ we get:
\begin{equation} \label{fngn sum}
    \sum_{n = -N+1}^Nf_ng_n = 2N a(z)-z \dot{a}(z).
\end{equation}
Now we can calculate the trace of the Green's function:
\begin{equation} 
    \text{Tr}(G) 
          = \sum \frac{f_ng_n}{w}
          %&=  \frac{2Na(z)-z\dot{a}(z)}{a(z)(\frac{1}{z}-z)} \nonumber\\
          %&= \frac{\dot{a}(z)}{a(z)(1-\frac{1}{z^2})} + \frac{2N}{\frac{1}{z}-z} \nonumber \\
          %& = \frac{1}{1-\frac{1}{z^2}}\frac{\partial \ln[a(z)]}{\partial z} + \frac{2N}{\frac{1}{z}-z} \nonumber \\ \label{Trace of G}
          = \frac{\partial \log a(y)}{\partial y} + \frac{2N}{\frac{1}{z}-z} \label{Trace of G},
\end{equation}
where $y=z+\frac{1}{z}$. For a free particle, with $u_n=u$ and $t_n=1$, the Green's function $G^0$ is given by the equation
\begin{equation} \label{G0 in matrix form}
    \left[H^0+\left(z+\frac{1}{z}-u\right)\mathbb{1}\right]G^0 = \mathbb{1},
\end{equation}
where the nonzero matrix elements of $H^0$ are $H_{n,n-1}^0=H_{n,n+1}^0=-1$ and $H_{n,n}^0=u$. One can show that
\begin{equation} \label{Trace of G0}
    \text{Tr}\left(G^0\right) = \frac{2N}{\frac{1}{z}-z}.
\end{equation}
Subtracting Eqn. \eqref{Trace of G0} from Eqn. \eqref{Trace of G}, we find the following equation:
\begin{equation} \label{subtraction of trace}
    \text{Tr}\left(G-G^0\right) = \frac{\partial \log a(y)}{\partial y}.
\end{equation}
Substituting Eqn. \eqref{G in matrix form} and Eqn. \eqref{G0 in matrix form} into Eqn. \eqref{subtraction of trace} we get:
\begin{equation}
    \frac{\partial \log a(y)}{\partial y} =  \text{Tr}\left((H+y)^{-1}-(H^0+y)^{-1}\right) = \sum_{n=1}\frac{(-1)^n \text{Tr}\left(H^n-(H^0)^n\right)}{y^{n+1}},
\end{equation}
where $y=z+\frac{1}{z}$. Integrating left hand side of the above equation from $y$ to $\infty$ we get Eqn. \eqref{connceting a with potential}.
\subsection{Dirac equation}

We derive trace identities associated with a discretized version of the Dirac operator, which to the authors' knowledge, has not been previously derived.  We consider the equations of motion
\begin{align}
    -\eta_n \psi^{(1)}_n + \tau_{n+1} \psi^{(2)}_{n+1} + \chi_n \psi^{(2)}_{n} &= \omega \psi^{(1)}_n \label{diraceom1}\\
    \tau_{n} \psi^{(1)}_{n-1} + \chi_n \psi^{(1)}_{n} + \eta_n \psi^{(2)}_n  &= \omega \psi^{(2)}_n \label{diraceom2}
\end{align}
which is the component form of the equations of motion for the chiral Gross-Neveu model.  For the scattering problem, we assume that both potentials $\eta$ and $\chi$, as well as the hopping $\tau$ converge to a constant at infinity.  The spectrum of the scattering states can then be determined via the substitution $\psi^{(1)}_n = \alpha(z) z^n$ and $\psi^{(2)}_n = \beta(z) z^n$:
\begin{equation}
    \begin{pmatrix} -\eta_{\infty} & \tau_{\infty} z + \chi_{\infty} \\ \tau_{\infty} z^{-1} + \chi_{\infty} & \eta_{\infty} \end{pmatrix} \begin{pmatrix} \alpha(z) \\ \beta(z)
    \end{pmatrix} = \omega \begin{pmatrix} \alpha(z) \\ \beta(z)
    \end{pmatrix}.
\end{equation}
The eigenvalues are given by
\begin{equation}
    \omega_{\pm}(\phi) = \pm \sqrt{\eta_{\infty}^2 + (\tau_{\infty} \cos \phi +\chi_{\infty})^2 + \tau_{\infty}^2 \sin^2\phi}.
\end{equation}
As before, we construct the Green's function in terms of the scattering states.  We first note that $\omega_{\pm}(\phi) = \omega_{\pm}(-\phi)$, so that each energy eigenstate is doubly degenerate.  We then construct the Jost functions $f_{\pm}$ and $g_{\pm}$ such that both functions mimic the asymptotic properties described in the main text.  In particular, we write that
\begin{equation}
    \lim_{n \to \infty} f_{\pm, n}(z) = z^n \begin{pmatrix}
    \alpha_1^{\pm}(z) \\ 
    \beta_1^{\pm}(z)
    \end{pmatrix}
\end{equation}
and
\begin{equation}
    \lim_{n \to -\infty} g_{\pm, n}(z) = z^{-n} \begin{pmatrix}
    \alpha_2^{\pm}(z) \\ 
    \beta_2^{\pm}(z)
    \end{pmatrix}
\end{equation}
We also have the following relation between the Jost functions, through which the transmission and reflection coefficients for the positive and negative energy modes are defined:
\begin{align}
    f_{\pm, n}(z) &= b_{\pm}(z) g_{\pm, n}(z) + a_{\pm}(z) g_{\pm, n}\left(z^{-1}\right),\\\
    g_{\pm, n}(z) &= \overline{b_{\pm}}(z) f_{\pm, n}(z) + \overline{a_{\pm}}(z) f_{\pm, n}\left(z^{-1}\right).
\end{align}
It can be seen that $\overline{a_{\pm}}(z) = a_{\pm}(z)$ and $\overline{b_{\pm}}(z) = -b_{\pm}\left(z^{-1}\right)$.  Then, we prove conservation of the Wronskian.  It can be shown by simple manipulation using the equations of motion that the quantity
\begin{equation}\label{diracwrons}
    w_n = \tau_{n+1}\left(f_n^{(1)} g_{n+1}^{(2)} - g_n^{(1)} f_{n+1}^{(2)}\right)
\end{equation}
is independent of $n$.  Utilizing the properties of the scattering states $f$ and $g$, the value of the Wronskian can be shown to equal
\begin{align}
    w_n &= \lim_{n \to -\infty} \tau_{n+1}\left(f_n^{(1)} g_{n+1}^{(2)} - g_n^{(1)} f_{n+1}^{(2)}\right)\\
    &= a(z) \alpha_2(z) \beta_2(z)\left(\frac{1}{z}-z\right),
\end{align}
where $\alpha_2$ and $\beta_2$ are the components of the eigenvector $g$. 

Next, we construct the Green's function, which now is comprised of four matrices, and is a solution to the equation
\begin{equation}\label{diracgreens}
    \begin{pmatrix} -\eta_n(t) - \omega(z) & \partial_{n}+\chi_n(t) \\ \partial_{n}^{\dagger}+\chi_{n}(t) & \eta_n(t) - \omega(z) \end{pmatrix} \begin{pmatrix} G_{11}(z) & G_{12}(z) \\ G_{21}(z) & G_{22}(z) \end{pmatrix} =  \begin{pmatrix} \mathbb{1} & 0 \\ 0 & \mathbb{1} \end{pmatrix}.
\end{equation}
It can be shown that the form of the Green's function matrix is
\begin{equation}
    \begin{pmatrix} G_{11}(z) & G_{12}(z) \\ G_{21}(z) & G_{22}(z) \end{pmatrix} = \frac{1}{w} \times \begin{cases}
     \begin{pmatrix} f_n^{(1)} g_m^{(1)} & f_n^{(1)} g_m^{(2)} \\ f_n^{(2)} g_m^{(1)} & f_n^{(2)} g_m^{(2)} \end{pmatrix} & \text{for } n > m \vspace{0.25cm} \\
     \begin{pmatrix} g_n^{(1)} f_m^{(1)} & g_n^{(1)} f_m^{(2)} \\ g_n^{(2)} f_m^{(1)} & g_n^{(2)} f_m^{(2)} \end{pmatrix} & \text{for } n < m \vspace{0.25cm} \\
     \begin{pmatrix} g_n^{(1)} f_m^{(1)} & f_n^{(1)} g_m^{(2)} \\ g_n^{(2)} f_m^{(1)} & f_n^{(2)} g_m^{(2)} \end{pmatrix} & \text{for } n = m
    \end{cases}
\end{equation}
where $w$ is the Wronskian.  This can be verified by direct substitution into Eqn. \eqref{diracgreens} as well as using the form of the Wronskian in Eqn. \eqref{diracwrons}.  We want to relate the trace of the Green's function to the values of the onsite potential and hopping:
\begin{equation}
    \text{Tr}(G) = \frac{1}{w} \sum_{n} (f_n^{(1)} g_n^{(1)} + f_n^{(2)} g_n^{(2)}).
\end{equation}
Next, we return to the equations of motion in Eqns. \eqref{diraceom1} and \eqref{diraceom2}, and take a derivative with respect to $z$
\begin{align}
    -\eta_n \dot{f}^{(1)}_n + \tau_{n+1} \dot{f}^{(2)}_{n+1} + \chi_n \dot{f}^{(2)}_{n} &= \dot{\omega} f^{(1)}_n + \omega \dot{f}^{(1)}_n\\
    \tau_{n} \dot{f}^{(1)}_{n-1} + \chi_n \dot{f}^{(1)}_{n} + \eta_n \dot{f}^{(2)}_n  &= \dot{\omega} f^{(2)}_n + \omega \dot{f}^{(2)}_n.
\end{align}
Multiplying the first equation by $g_n^{(1)}$ and the second by $g_n^{(2)}$ gives
\begin{align}
    -\eta_n g_n^{(1)}\dot{f}^{(1)}_n + \tau_{n+1} g_n^{(1)}\dot{f}^{(2)}_{n+1} + \chi_n g_n^{(1)}\dot{f}^{(2)}_{n} &= \dot{\omega} g_n^{(1)}f^{(1)}_n + \omega g_n^{(1)} \dot{f}^{(1)}_n\\
    \tau_{n} g_n^{(2)}\dot{f}^{(1)}_{n-1} + \chi_n g_n^{(2)}\dot{f}^{(1)}_{n} + \eta_n g_n^{(2)}\dot{f}^{(2)}_n  &= \dot{\omega} g_n^{(2)}f^{(2)}_n + \omega g_n^{(2)}\dot{f}^{(2)}_n.
\end{align}
Next, we return to the original equations of motion now for $g$; multiplying the first equation of motion by $\dot{f}_n^{(1)}$ and the second equation of motion by $\dot{f}_n^{(2)}$, we find
\begin{align}
    -\eta_n \dot{f}_n^{(1)} g^{(1)}_n + \tau_{n+1} \dot{f}_n^{(1)} g^{(2)}_{n+1} + \chi_n \dot{f}_n^{(1)} g^{(2)}_{n} &= \omega \dot{f}_n^{(1)} g^{(1)}_n\\
    \tau_{n} \dot{f}_n^{(2)} g^{(1)}_{n-1} + \chi_n \dot{f}_n^{(2)} g^{(1)}_{n} + \eta_n \dot{f}_n^{(2)} g^{(2)}_n  &= \omega \dot{f}_n^{(2)} g^{(2)}_n
\end{align}
Combining these equations, we find that
\begin{align}
    \tau_{n+1} \left(g_n^{(1)}\dot{f}^{(2)}_{n+1} - \dot{f}_n^{(1)} g^{(2)}_{n+1}\right) + \chi_n \left(g_n^{(1)}\dot{f}^{(2)}_{n} - \dot{f}_n^{(1)} g^{(2)}_{n}\right) &= \dot{\omega} g_n^{(1)}f^{(1)}_n\\
    \tau_{n}\left(g_n^{(2)}\dot{f}^{(1)}_{n-1} - \dot{f}_n^{(2)} g^{(1)}_{n-1}\right) + \chi_n\left(g_n^{(2)}\dot{f}^{(1)}_{n}-\dot{f}_n^{(2)} g^{(1)}_{n}\right) &= \dot{\omega} g_n^{(2)}f^{(2)}_n
\end{align}
Therefore, adding both equations gives us
\begin{equation}
    \dot{\omega} \left(g_n^{(1)}f^{(1)}_n + g_n^{(2)}f^{(2)}_n\right) = U_{n+1} - U_{n},
\end{equation}
where
\begin{equation}
   U_{n} = \tau_{n} \left(g_{n-1}^{(1)}\dot{f}^{(2)}_{n} - \dot{f}_{n-1}^{(1)} g^{(2)}_{n}\right).
\end{equation}
Taking a sum over $n$, we obtain the expression
\begin{equation}
    \dot{\omega} \sum_n \left(g_n^{(1)}f^{(1)}_n - g_n^{(2)}f^{(2)}_n\right) = U_{\infty} - U_{-\infty}.
\end{equation}
We can compute the asymptotic form of $U_n$ at $n \to \infty$, yielding
\begin{align}
    \lim_{n \to \infty} U_n &= \tau_{\infty} a(z) \left(z^{-n+1} \alpha_1(z) \frac{d}{dz}(\beta_1(z) z^n) - z^{-n} \beta_1(z) \frac{d}{dz}(\alpha_1(z) z^{n-1})\right) \\
&= \tau_{\infty} a(z) \left(\alpha_1 \dot{\beta}_1 z - \beta_1 \dot{\alpha}_1 z^{-1} + \alpha_1 \beta_1\left(n - (n-1)z^{-2}\right)\right).
\end{align}
To compute $U_n$ in the opposite limit, we utilize the identity
\begin{equation}
   U_{n} = \frac{d}{dz} w(z) - \tau_{n} \left(\dot{g}_{n-1}^{(1)}f^{(2)}_{n} - f_{n-1}^{(1)} \dot{g}^{(2)}_{n}\right),
\end{equation}
which therefore gives
\begin{align}
   \lim_{n \to -\infty}U_{n} &= \frac{d}{dz} w(z) - \tau_{-\infty} a(z)\left(z^n \beta_2(z) \frac{d}{dz}\left(z^{-n+1} \alpha_2(z)\right) - z^{n-1} \alpha_2(z) \frac{d}{dz}\left(z^{-n} \beta_2(z)\right)\right) \\
   &= \frac{d}{dz} w(z) - \tau_{-\infty} a(z)\left(\dot{\alpha}_2(z) \beta_2(z) z - \alpha_2(z) \dot{\beta}_2(z) z^{-1} + \alpha_2(z) \beta_2(z)\left(n z^{-2} - (n-1)\right)\right).
\end{align}
With these asymptotic forms defined, it is important to note that their difference will be a function of $a(z)$ and $\dot{a}(z)$, but the derivative contribution comes solely from the derivative of the Wronskian.  Then, we may write the Green's function as
\begin{equation}
    \text{Tr}(G) = \frac{U_{\infty} - U_{-\infty}}{\dot{\omega} w} = -\frac{\dot{w}}{\dot{\omega} w} + \frac{\xi(z)}{\dot{\omega}},
\end{equation}
where $\xi(z)$ is a known function of $z$ and can be computed based on the asymptotic components of the Jost functions.  Next, we compute the trace of the Green's function by taking the matrix inverse of the relation in Eqn. \eqref{diracgreens}.  Splitting the matrix into $H - \omega(z) \mathbb{1}$ and performing a series expansion about $z = 0$ or $\omega(z) \to \infty$, we find that
\begin{equation}
    \text{Tr}(G) = -\frac{1}{\omega(z)}\sum_{k = 0}^\infty \frac{\text{Tr}(H^k)}{\omega^k(z)},
\end{equation}
Combining this with the previous equation and integrating with respect to $z$, we find
\begin{equation}
   -\log \frac{w(z)}{w(0)} + \int_0^z \xi(z') \,dz = \sum_{k = 0}^\infty \frac{\text{Tr}(H^{2k})}{\omega^{2k} k},
\end{equation}
Using the explicit form of the Wronskian, the logarithm of the reflection coefficient can be written as
\begin{equation}\label{notlaurent}
   -\log \frac{a(z)}{a(0)} = h(z) + \sum_{k = 0}^\infty \frac{\text{Tr}(H^{2k})}{\omega^{2k} k},
\end{equation}
where $h(z)$ is a known albeit complicated function of $z$.  It admits a series expansion in $z$, but the coefficients are constants and only depend on asymptotics of $\Delta$, $\tau$, and $\Pi$.  This contribution, although infinitely large, can be ignored by a suitable shift of the vacuum energy in the action.  It is also important to note that the analysis above holds for only one of the energy branches, which we choose to be the positive one.  Notice that the series expansion on the right hand side of Eqn. \eqref{notlaurent} is not a Laurent series; for odd powers of $\omega$, the series expansion involves half-integer powers of $z$.  Thus, we cannot claim that $\log a_+$ or $\log a_-$ individually is meromorphic within the unit disk.  However, if we add the positive and negative branches, then we find
\begin{align}
   -\log \frac{a_+(z)}{a_+(0)} - \log \frac{a_-(z)}{a_-(0)} &= h_+(z) + h_-(z) + \sum_{k = 0}^\infty \frac{\text{Tr}(H^k)}{\omega^k k} + \sum_{k = 0}^\infty \frac{(-1)^k \text{Tr}(H^k)}{\omega^k k} \\
   &= h_+(z) + h_-(z) + \sum_{k = 0}^\infty \frac{\text{Tr}(H^{2k})}{\omega^{2k} k}.
\end{align}
As even powers of $\omega$ admit a Laurent series expansion, the left hand side is now meromorphic in the unit disk and the Poisson-Jensen formula may be applied.  The application of the Poisson-Jensen formula is discussed in the text.

\section{Relating Floquet indices to phase shift
}
\label{Appendix B}

In this appendix we derive Eqn. \eqref{appendixb}, which gives a relation between the Floquet indices and the phase shift of the scattering problem.  We want to compute $\sum_{k} |\alpha_k|$, where $\alpha_k = \sqrt{u-2\cos \phi_k)}$.  This term is normally divergent; thus to tame the divergence and extract the finite contribution, we place the system in a finite box of lattice spacing $N$ and impose the periodic boundary conditions $\psi_i(t) = \psi_{i+N}(t)$ before taking the limit $N \to \infty$.  Next, we construct the scattering problem for the potential $u_n^{(1)}$ and hoppings $t_n^{(1)}$.  In terms of left and right movers, we may write the scattering wavefunction as
\begin{equation}
    \begin{pmatrix} \Psi_{L,n} \\ \Psi_{R,n} \end{pmatrix} = \begin{pmatrix} e^{i n \phi} \text{ as } n \to -\infty \\ e^{-i n \phi} \text{ as } n \to \infty \end{pmatrix} + S \begin{pmatrix} e^{i n \phi} \text{ as } n \to \infty \\ e^{-i n \phi} \text{ as } n \to -\infty \end{pmatrix},
\end{equation}
where $S$ is the scattering matrix for the one dimensional scattering problem.  In terms of the transmission and reflection amplitudes, $S$ can be written as
\begin{equation}
    S = \begin{pmatrix} t & r \\ -r^* \frac{t}{t^*} & t \end{pmatrix},
\end{equation}
where $t$ and $r$ are the transmission and reflection coefficients, respectively.  As $S$ is unitary, it is diagonalized by some matrix $M$, whose columns are the eigenvectors of $S$.  As a result, we may write
\begin{align}
    M \begin{pmatrix} \Psi_{L,n} \\ \Psi_{R,n} \end{pmatrix} = M &\begin{pmatrix} e^{i n \phi} \text{ as } n \to -\infty \\ e^{-i n \phi} \text{ as } n \to \infty \end{pmatrix} + \Lambda M \begin{pmatrix} e^{i n \phi} \text{ as } n \to \infty \\ e^{-i n \phi} \text{ as } n \to -\infty \end{pmatrix},
\end{align}
where $\Lambda$ is a diagonal matrix whose elements are the eigenvalues of $S$, which we label as $\exp(i \lambda_\pm)$.  The functions defined on the left which we will rename $\overline{\Psi}_L$ and $\overline{\Psi}_R$ will be basis functions for solutions that satisfy periodic boundary conditions.  In particular, setting $\overline{\Psi}_{L,-N/2} = \overline{\Psi}_{L,N/2}$ and $\overline{\Psi}_{R,-N/2} = \overline{\Psi}_{R,N/2}$ gives the equations $\exp\left(i N \phi + i \lambda_{\pm}\right) = 1$.  This results in two branches of solutions for the quasi-momentum $\phi$:
\begin{equation}\label{phibranch}
    \phi_{\pm, n} = \frac{2 \pi n - \lambda_{\pm}}{N}.
\end{equation}
Next, we return to the original expression
\begin{align}
    \sum_k |\alpha_k| = T \sum_{s = \pm, n} \sqrt{u - 2 \cos \phi_{s, n}}
\end{align}
Using Eqn \ref{phibranch} and expanding in a Taylor series of $1/N$, we obtain up to $O(1/N)$
\begin{align}
    \sum_k |\alpha_k| = T \sum_{s, n} \sqrt{u - 2 \cos \frac{2 \pi n}{N}} + T \sum_{s, n} \frac{\lambda_s}{N} \frac{\sin \frac{2 \pi n}{N}}{\sqrt{u - 2 \cos \frac{2 \pi n}{N}}}.
\end{align}
The first sum is a divergent constant due to the lattice regularization scheme that we use.  The finite part in the limit $N \to \infty$ will be
\begin{equation}
    \sum_k |\alpha_k| \stackrel{\text{finite}}{=} T \int_0^{2\pi} \frac{d\phi}{2 \pi} \frac{\sin \phi}{\sqrt{u - 2 \cos \phi}} \left[\lambda_+(\phi) + \lambda_-(\phi)\right].
\end{equation}
The eigenvalues of $S$ can be readily computed and we find that the two eigenvalues are $\lambda_\pm = \delta_t(\phi) \pm \tan^{-1} |r(\phi)/t(\phi)|$ where $\delta_t(\phi)$ is the phase of the transmission coefficient.  Therefore, we find
\begin{align}
    \sum_k |\alpha_k| = 2T \int_0^{2\pi} \frac{d\phi}{2 \pi} \frac{\sin \phi}{\sqrt{u - 2 \cos \phi}} \delta_t(\phi).
\end{align}
A nearly identical analysis holds in the case of the chiral Gross-Neveu model; here, we have two pairs of scattering problems that are decoupled, so one must separately add the phase of the transmission coefficients for both of the scattering problems in the right hand side of the equation above.

\acknowledgments

This work was supported by the U.S. Department of Energy, Office of Science, Basic Energy Sciences under Award No. DE-SC0001911 and the Simons Foundation.  S.B. was additionally supported by the National Science Foundation Graduate Research Fellowship under Grant No. 1745302.  

\paragraph{Note added.} After completion of this manuscript, the authors became aware of \cite{SPIRIDONOV1995126}, which focuses on constructing exact hierarchies of discrete reflectionless potentials using a method similar to supersymmetric quantum mechanics (albeit more restrictive than ours).  It would be interesting to understand whether more kinds of exact hierarchies can be constructed using our framework.  

\bibliography{references}
\bibliographystyle{unsrt}
%\begin{thebibliography}{99}

%\bibitem{a}
%Author, \emph{Title}, \emph{J. Abbrev.} {\bf vol} (year) pg.

%\bibitem{b}
%Author, \emph{Title},
%arxiv:1234.5678.

%\bibitem{c}
%Author, \emph{Title},
%Publisher (year).

%\end{thebibliography}
\end{document}